\PassOptionsToPackage{table,xcdraw}{xcolor}
\documentclass[sigconf,screen]{acmart}
\usepackage{graphicx}
\usepackage{subfigure}
\usepackage{comment}
\usepackage{color}
\usepackage{amsmath}
\usepackage{mathtools}
\usepackage{fixltx2e}
\usepackage{bbm}
\usepackage{pifont}
\usepackage{booktabs} 
\usepackage{multirow}
\usepackage{algorithm}
\usepackage{array}
\usepackage{enumitem}
\usepackage{amsthm}
\usepackage{pifont}
\usepackage{arydshln}
\DeclareMathOperator*{\argmax}{argmax}

\usepackage[noend]{algpseudocode}
\algdef{SE}[DOWHILE]{Do}{doWhile}{\algorithmicdo}[1]{\algorithmicwhile\ #1}%
\PassOptionsToPackage{hyphens}{url}
\usepackage[export]{adjustbox}
\usepackage{multicol}
\usepackage{tikz}

\theoremstyle{definition}

\definecolor{darkgray}{rgb}{0.3,0.3,0.3}
\definecolor{tblue}{rgb}{0.09,0.35,0.66}

\newcommand{\setred}[1]{{\color{darkred}{#1}}}
\newcommand{\setblue}[1]{{\color{tblue}{#1}}}

\definecolor{mygray}{gray}{.9}

\copyrightyear{2023}
\acmYear{2023}
\setcopyright{rightsretained}
\acmConference[MM '23]{Proceedings of the 31st ACM International Conference
on Multimedia}{October 29-November 3, 2023}{Ottawa, ON, Canada}
\acmBooktitle{Proceedings of the 31st ACM International Conference on
Multimedia (MM '23), October 29-November 3, 2023, Ottawa, ON, Canada}
\acmDOI{10.1145/3581783.3611771}
\acmISBN{979-8-4007-0108-5/23/10}

\usepackage{etoolbox}
\makeatletter
\patchcmd{\maketitle}{\@copyrightpermission}{
   \begin{minipage}{0.3\columnwidth}
     \href{https://creativecommons.org/licenses/by/4.0/}{\includegraphics[width=0.90\textwidth]{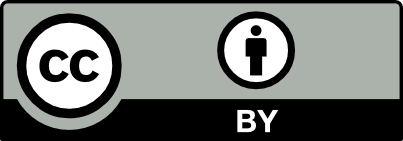}}
   \end{minipage}\hfill
   \begin{minipage}{0.7\columnwidth}
     \href{https://creativecommons.org/licenses/by/4.0/}{This work is licensed under a Creative Commons Attribution International 4.0 License.}
   \end{minipage}
  
   \vspace{5pt}
}{}{}

\makeatother
 
\usepackage{dutchcal}

\begin{document}
%
\title{Optimizing Adaptive Video Streaming with Human Feedback}


\author{Tianchi Huang}
\authornote{Both authors contributed equally to this research.}
\authornote{Corresponding Author}
\affiliation{%
  \institution{Tsinghua University}
}
\email{htc19@mails.tsinghua.edu.cn}

\author{Rui-Xiao Zhang}
\authornotemark[1]
\affiliation{%
  \institution{The University of Hong Kong}
}
\email{zrxhku@hku.hk}

\author{Chenglei Wu}
\affiliation{%
  \institution{Tencent}
}
\email{clwwwu@tencent.com}

\author{Lifeng Sun}
\authornotemark[2]
\affiliation{%
  \institution{BNRist, Tsinghua University}
}
\email{sunlf@tsinghua.edu.cn}

\begin{abstract}
Quality of Experience~(QoE)-driven adaptive bitrate~(ABR) algorithms are typically optimized using QoE models that are based on the mean opinion score~(MOS), while such principles may not account for user heterogeneity on rating scales, resulting in unexpected behaviors. In this paper, we propose \texttt{Jade}, which leverages reinforcement learning with human feedback~(RLHF) technologies to better align the users' opinion scores. \texttt{Jade}'s rank-based QoE model considers relative values of user ratings to interpret the subjective perception of video sessions. We implement linear-based and Deep Neural Network (DNN)-based architectures for satisfying both accuracy and generalization ability. We further propose entropy-aware reinforced mechanisms for training policies with the integration of the proposed QoE models. Experimental results demonstrate that \texttt{Jade} performs favorably on conventional metrics, such as quality and stall ratio, and improves QoE by 8.09\%-38.13\% in different network conditions, emphasizing the importance of user heterogeneity in QoE modeling and the potential of combining linear-based and DNN-based models for performance improvement.
\end{abstract}

\keywords{Adaptive Video Streaming, Reinforcement Learning.}

\ccsdesc[300]{Information systems~Multimedia streaming}

\maketitle
\begin{figure*}
    \centering
    \subfigure[Ratings on Different Videos]{
        \includegraphics[width=0.23\linewidth]{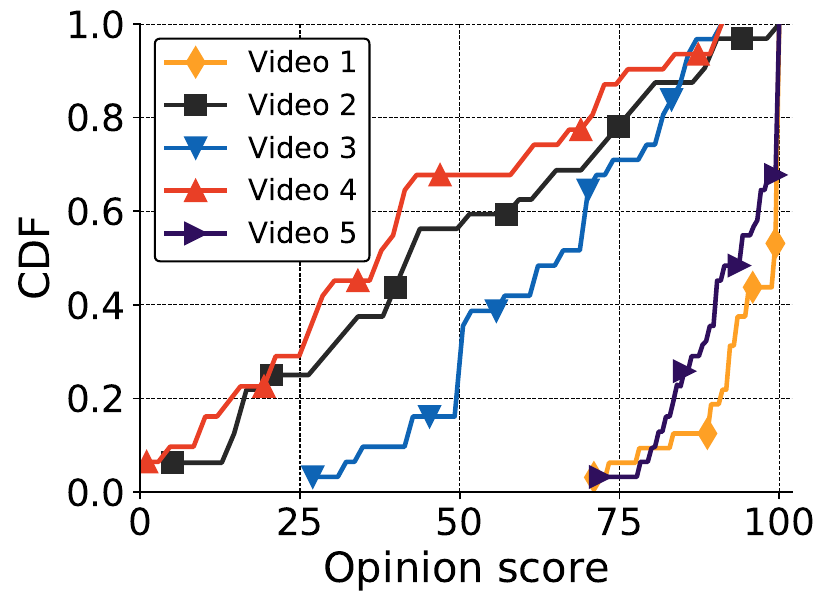}
        \label{fig:t1}
    }
    \subfigure[Distribution of Ratings for Videos]{
        \includegraphics[width=0.23\linewidth]{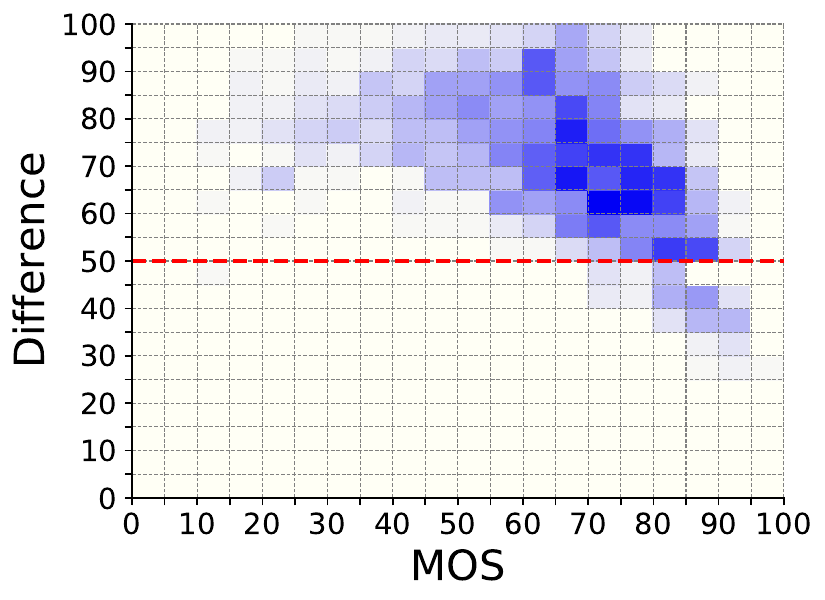}
        \label{fig:t2}
    }
    \subfigure[User Heterogeneity on Video Set]{
        \includegraphics[width=0.23\linewidth]{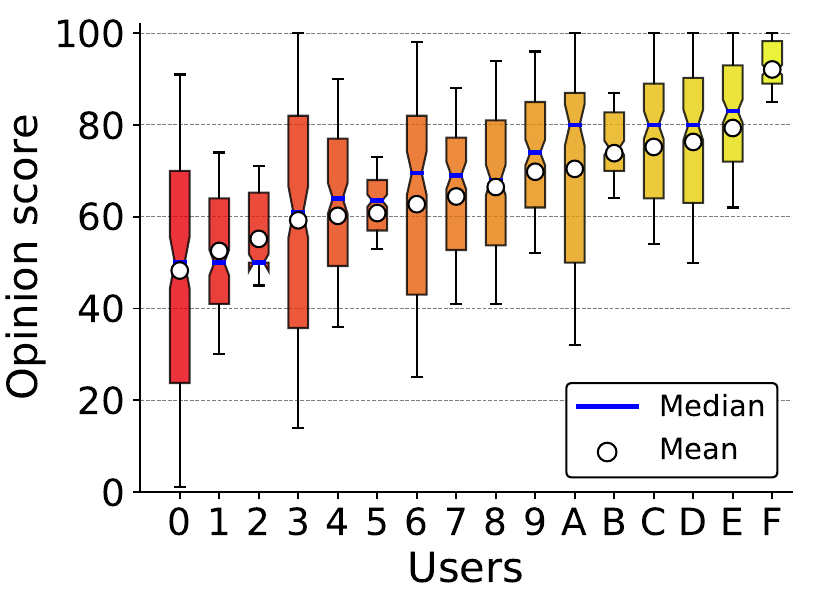}
        \label{fig:t3}
    }
    \subfigure[Uniformity in Criteria for Rating]{
        \includegraphics[width=0.23\linewidth]{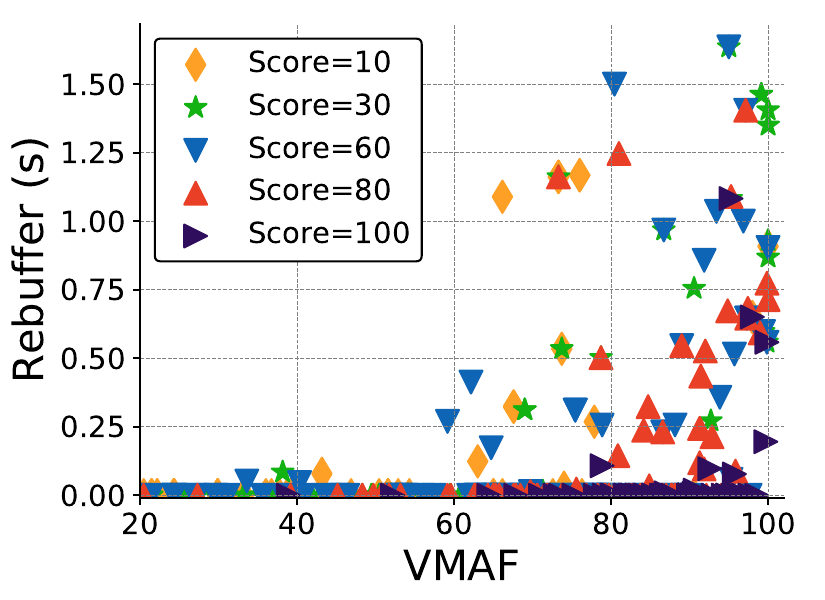}
        \label{fig:t4}
    }
    \vspace{-15pt}
    \caption{Visualizing Cumulative Distribution Function (CDF) of videos' opinion scores, the relationship between MOS and its variance of different videos, and user heterogeneity over the SQoE-IV dataset~\cite{duanmu2020assessing}.}
    \vspace{-10pt}
    \label{fig:motivation}
\end{figure*}

\section{Introduction}
In recent years, video services have become an integral part of people's daily lives due to the rapid proliferation of network technologies like 5G and the increasing demand for content expression by users. According to \emph{The 2022 Global Internet Phenomena Report}~\cite{SANDVINE2022}, video traffic accounted for a substantial share of 65.93\% of total internet traffic in the first half of 2022, showing a 24\% increase compared to the same period in 2021. Among them, adaptive video streaming has become one of the mainstream technologies for network video traffic, accounting for 40.24\% of the total traffic.

Recent adaptive bitrate (ABR) algorithms can be categorized into two types: Quality of Service (QoS)-driven approaches and Quality of Experience (QoE) optimization methods~(\S\ref{sec:related}). QoS-driven approaches focus on observed metrics and aims to achieve higher bitrate while avoiding rebuffering events through techniques such as throughput prediction~\cite{jiang2014improving,sun2016cs2p,lv2022lumos} and buffer control~\cite{huang2015buffer,yadav2017quetra,spiteri2016bola,spiteri2020bola}. However, despite their efficiency, such heuristics often fall short of fulfilling users' QoE requirements. As a result, various attempts have been made to directly optimize QoE ~\cite{yan2020learning,mao2017neural,huang2018qarc,huang2020stick}, in which the QoE model is predefined~\cite{yin2015control,bentaleb2016sdndash,li2023mamba} or learned from existing QoE datasets based on mean opinion scores~(MOS)~\cite{huang2020quality,li2022apprenticeship,turkkan2022greenabr}.

Unfortunately, our analysis on the SQoE-IV assessment database~\cite{duanmu2020assessing}, which includes ratings from 32 discerning users, reveals significant diversity in opinion scores and rating scales. Such findings challenge the effectiveness of MOS-based QoE models and emphasize the subjective nature of user perception~(\S\ref{sec:keyfindings}). Therefore, we consider a novel methodology to mitigate the bias in human feedback induced by user heterogeneity, aiming to achieve the best possible performance for all users.

We propose \texttt{Jade}, which leverages Reinforcement Learning with Human Feedback (RLHF) and a rank-based QoE model to learn a Neural Network~(NN)-based ABR algorithm~(\S\ref{sec:jade}). Unlike previous approaches, we consider relative values of user ratings to interpret the subjective perception of video sessions. Based on the perceived users' feedback, \texttt{Jade} solves two key challenges, i.e., train a rank-based QoE model~(\S\ref{sec:pairwise}), and design a deep reinforcement learning~(DRL)-based method for ABR algorithm generation~(\S\ref{sec:rlhf}).

Unlike previous MOS-based QoE methods~\cite{huang2019comyco,turkkan2022greenabr}, we propose a rank-based QoE model for video session ranking. Our key idea is to learn a ``ranked function'' aligned with the rank of each user's opinion score. Modeling as a Deep Structured Semantic Model (DSSM)~\cite{huang2013learning}, we implement a pairwise training methodology and learn two NN architectures, i.e., linear-based and Deep NN~(DNN)-based, to tame the models' generalization ability~(\S\ref{sec:design}) -- results show that the DNN-based QoE model achieves the highest accuracy but it exhibits imperfections in generalization and stability, while the linear-based model performs the opposite~(\S\ref{sec:qoemodeleval}).

We present a DRL-based approach~\cite{ye2019mastering} to generate an NN-based ABR algorithm based on the trained QoE models. In detail, we incorporate two entropy-aware mechanisms, i.e., smooth training and online trace selection, to fully leverage the advantages of the QoE models. The smooth training methods leverage the linear-based QoE model in the early stage and transition to the DNN-based QoE model in the later stage, with an entropy-related parameter to combine their outputs for seamless integration, which allows for effective training and integration of the two QoE models during different training phases~(\S\ref{sec:smoothtraining}). The online trace selection scheme adopts online learning techniques to dynamically choose the appropriate network trace for training~(\S\ref{sec:OnlineTraceSelection}). 

We evaluate \texttt{Jade} against recent ABRs in both slow-network and fast-network paths using trace-driven simulation~(\S\ref{sec:slowfast}).
In slow-network paths, \texttt{Jade} outperforms existing algorithms by 22.5\%-38.13\%, including RobustMPC and Pensieve. Compared with Comyco, \texttt{Jade} has a relatively improved QoE of 8.09\%. \texttt{Jade} achieves the Pareto Frontier, indicating optimal trade-offs between different objectives.
In fast-network paths, \texttt{Jade} improves QoE by at least 23.07\% compared to other QoE-driven ABR algorithms, highlighting the importance of considering both the ``thermodynamics'' and ``kinetics'' of the process in QoE modeling.
Further experiments demonstrate the potential risks of relying on ``imperfect QoE models'' without proper caution, as it may result in misguided strategies and undesired performances. We prove that through the synergistic fusion of linear-based and DNN-based QoE models, \texttt{Jade} outperforms approaches that utilize linear-based or DNN-based models solely~(\S\ref{sec:jadevsqoe}). We sweep the parameter settings of the QoE model comparison and validate the effectiveness of smooth training and online trace selector in enhancing performance and accelerating the learning process by selecting useful traces~(\S\ref{sec:abl}).

In general, we summarize the contributions as follows:

\begin{itemize}[leftmargin=*]
    \item We identify limitations in recent QoE models, particularly the use of MOS which lacks consideration for user heterogeneity, leading to sub-optimal policies in the learned ABR algorithms~(\S\ref{sec:background}).
    \item We propose \texttt{Jade}, the first RLHF-based ABR system synthesizes human feedback to develop an NN-based ABR algorithm, including a rank-based QoE model and an entropy-aware DRL-based learning process~(\S\ref{sec:jade}, \S\ref{sec:pairwise}, \S\ref{sec:rlhf}).
    \item We implement Jade and evaluate its performance across various network conditions. Results demonstrate its superior QoE performance in diverse scenarios~(\S\ref{sec:eval}).
\end{itemize}
\vspace{-10pt}
\section{Background and Motivation}
\label{sec:background}
\subsection{Adaptive Video Streaming}
The conventional adaptive video streaming framework consists of a video player with a limited buffer length~(i.e., typically 40 to 240 seconds) and an HTTP-Server or Content Delivery Network~(CDN)~\cite{bentaleb2018survey}.
On the server-side, raw videos are chunked into segments, typically lasting 2 to 10 seconds~\cite{yan2020learning,mao2017neural}. These segments are then encoded at different bitrates or quality levels before being stored on the designated storage server~\cite{huang2021deep}. 
Adaptive Bitrate (ABR) algorithms, such as HTTP Live Streaming~(HLS)\cite{HLS} and DASH~\cite{dash} utilize throughput estimation~\cite{jiang2014improving} and playback buffer occupancy~\cite{huang2015buffer} to determine the most appropriate bitrate level.

\subsection{Key findings}
\label{sec:keyfindings}
Off-the-shelf QoE for adaptive video streaming is calculated by the arithmetic mean of subjective judgments using a 5-point absolute category rating (ACR) quality scale~\cite{rec2006p}, or leveraging a continuous scale ranging between 1–100~\cite{huynh2010study}.
However, such schemes fail to consider the effect of user heterogeneity over opinion scores.
To this end, we ask: \emph{does recent mean opinion score~(MOS)-based QoE model work on the right track?} 
If it does work, the following conditions should be satisfied: i) the user's opinion score for each video session should be consistent, and ii) the user should assign similar scores to similar video sessions.

\textbf{Dataset.}~Recent publicly available QoE datasets either lack comprehensive feedback information~\cite{bampis2018towards,duanmu2018quality} or have not yet been made available for publication~(e.g., Ruyi~\cite{zuo2022adaptive}). 
We use SQoE-IV assessment database, which contains 1350 subjective-rated streaming video sessions that are generated from diverse video sources, video codecs, network conditions, ABR algorithms, as well as viewing devices~\cite{duanmu2020assessing}. 
Each video session is assessed by a panel of 32 users, who provide their discerning ratings. Thus, the dataset totally consists of 43,200 opinion scores, which is sufficient for data analysis.

\textbf{Video-wise analysis.}~We start by investigating the correlation between videos and opinion scores, where the videos are the sessions being performed by different ABR algorithms.
Figure~\ref{fig:t1} shows the CDF of opinion score over different videos, where the videos are randomly picked from the SQoE-IV dataset. Notably, diverse opinion scores are observed for the same videos, with video 1 exhibiting scores ranging from 60 to 100, and other videos (i.e., video 2 and video 4) even displaying scores ranging from 0 to 100, encompassing the full range of minimum to maximum opinion scores. These findings implicitly highlight that users hold varying opinions about the same video, albeit with a variance that may exceed our initial expectations.
Figure~\ref{fig:t2} illustrates the detailed proportion occupied by ranges, where the difference metric represents the gap between the maximum and the minimum opinion score over each video. We observe that almost all the videos are scored with higher differences, as few video result in a difference of below 50. Hence, the user's rating of each session is diverse rather than consistent. Such high-variance results also motivate us to further explore the key reasons from the user-level analysis.

\textbf{User Heterogeneity.}
We report the box plot of 16 users' feedback in Figure~\ref{fig:t3}, here all the results are sorted and collected from the same video set. Surprisingly, we find diverse rating scales on the user level. For instance, some users rate the opinion score from 0~(user 0) while others even attempt to start their evaluation process with 60~(user E). At the same time, we also observe the diverse preference of different users. User F prefers rating higher scores over all sessions but most users are conservative, as the average score is often lower than the median score over all sessions. 
Detailing users' opinions in Figure~\ref{fig:t4}, we observe that there is a lack of uniformity in the criteria used by each user to assign the same score -- given the session with the same video quality~(i.e., VMAF~\cite{rassool2017vmaf}) and rebuffering time~(i.e., the most essential metrics to evaluate QoE~\cite{duanmu2018quality}), users provide varying feedback. 
When the video quality reaches 70 and the rebuffering time exceeds 1.0 seconds, the rated scores provided by users exhibit heterogeneous, spanning from 10 to 80.
Thus, we argue that users assign diverse opinion scores for the same session. In other words, the perception of the same score, such as 80, may vary among different users.

To sum up, the expected results, as indicated by the conditions of consistent opinion scores from users for each video session and similarity in scores assigned to similar video sessions, have not been achieved w.r.t the observed results.
To that end, despite the outstanding performance of recent work, such MOS-based QoE metrics may not perform on the right track.

\begin{figure}
    \centering
    \includegraphics[width=1.0\linewidth]{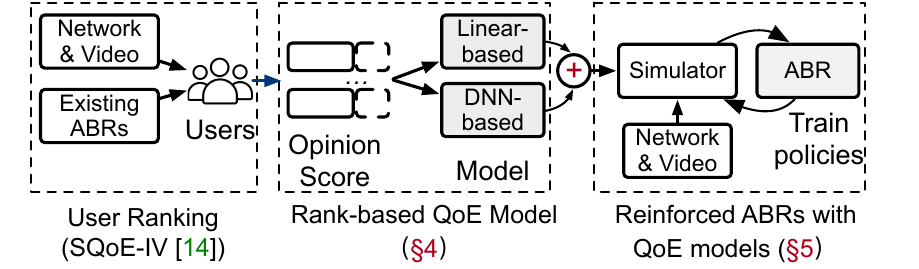}
    \vspace{-20pt}
    \caption{\texttt{Jade}'s system overview. Based on users' feedback of ranking scores, \texttt{Jade} trains rank-based QoE models for learning an NN-based ABR algorithm.}
    \vspace{-15pt}
    \label{fig:bigpicture}
\end{figure}

\section{\texttt{Jade} Overview}
\label{sec:jade}
In this work, we utilize Reinforcement Learning with Human Feedback (RLHF)~\cite{bai2022training} to address the problem at hand by leveraging the rank-based QoE model as a reward signal.
Our key idea is to leverage a ranking-based method that naively ignores \emph{absolute values} but only focuses on the \emph{relative values} of scores assigned by each user -- if a user rates video session A higher than video session B, we interpret this as indicating a superior experience for A compared to B, according to the user's subjective perception. Unlike recent work, we require a rank-based QoE model to accurately generates ``virtual values'' that align with the relative ranking of users' ratings. By leveraging the trained QoE model as a reward signal, we can generate NN-based ABR algorithms with state-of-the-art Deep Reinforcement Learning~(DRL)~\cite{schulman2017proximal,ye2019mastering} methods, as this problem inherently falls within the purview of DRL.

We propose \texttt{Jade}, which can be viewed as the first RLHF-based ABR algorithm to our best knowledge. 
The big picture of \texttt{Jade} is illustrated in Figure~\ref{fig:bigpicture}. As shown, \texttt{Jade} is mainly composed of three phases, i.e., user rating, learning rank-based QoE models, and generating ABR algorithms with the learned QoE models.

The \texttt{Jade}'s workflow is shown as follows. \ding{182} The user rates opinion scores for the video sessions with different ABR algorithms, network traces, and video descriptions. In this work, we directly adopt the SQoE-IV dataset, consisting of 1350 realistic streaming videos generated from various transmitters, channels, and receivers. 
\ding{183} In the rank-based QoE model phase~(\S\ref{sec:pairwise}), we leverage users' ``relative'' feedback as the input for implementing a more effective and accurate QoE model. Different from previous work, 
considering model generalization and precision, we train two types of QoE models, including linear-based and DNN-based. 
\ding{184} In the training phase, we propose an entropy-aware reinforcement learning method for generating an NN-based ABR algorithm with the guidance of the trained QoE model~(\S\ref{sec:rlhf}).

\section{Rank-based QoE Model}
\label{sec:pairwise}
In this section, we propose a practical rank-based QoE model that aligns users' feedback scores, design a pairwise loss to train models, and introduce a novel evaluation metric, the Identity Rate~\footnote{In psychology, \emph{identity} refers to an individual's unique traits and qualities that make up their personal and social identity~\cite{adams1996developmental}.}, to validate the proposed QoE model.

\begin{figure}
    \centering
    \includegraphics[width=0.80\linewidth]{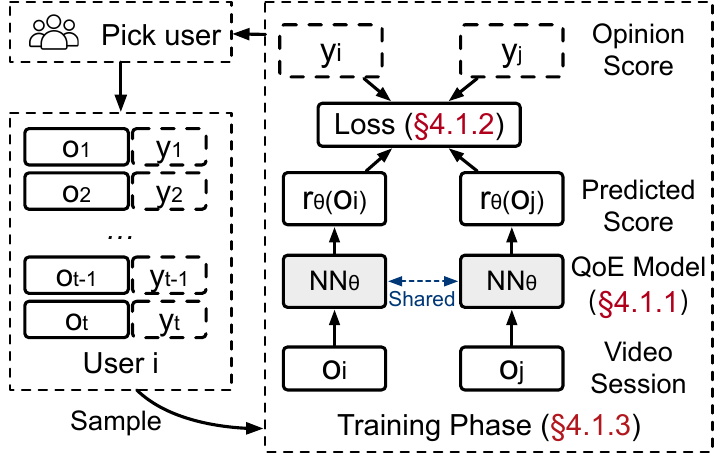}
    \vspace{-10pt}
    \caption{The training workflow of \texttt{Jade}'s rank-base QoE model, which utilizes pairwise loss to align the rank of the same user's opinion score over two sessions.}
    \vspace{-15pt}
    \label{fig:qoemodel}
\end{figure}

\subsection{Model Design}
\label{sec:design}
Different from prior work that integrates all the scores as the \emph{mean opinion score}~\cite{huang2019comyco,turkkan2022greenabr}, our key idea is to create a ``ranked function'' to map the opinion score for each video session, where the function is learned by pairs of samples in the queries that are previously fed back by the same user. Now we formally give a general description.

\begin{definition}
\emph{
We define a set of queries $Q=\{q^1, q^2, \dots, q^m\}$. 
For each query $q^i$, it is composed by a set of video sessions $O^i=\{o_1^i, o_2^i, \dots, o_t^i\}$. 
Each video session records video playback information for the entire session.
At the same time, for each user $u$, the session set $O^i$ is corresponding to a list of scores $Y_u^i=\{y_u^i(o_1^i), \dots, y_u^t(o_t^i)\}$, where $y_u^t(\cdot)$ reflects the opinion score judged by user $u$ according to the session $o^t$.
}
\end{definition}

In this paper, the opinion score $y_u^t$ ranges from 0 to 100. The higher the score, the higher the user's satisfaction with the video session. The playback information contains several underlying metrics such as video qualities, video bitrates, and rebuffering time for each video chunk.
Given a rating pair $\{y_u(o_i), y_u(o_j)\}$ over a user $u$, we further define a binary scalar $a_i^j$, which denotes the relative relationship between the two scores. 

\begin{small}
\begin{equation}
    a_i^j=\left\{
    \begin{array}{rcl}
    1 & & y(s_i) > y(s_j),\\
    -1 & & y(s_i) < y(s_j),\\
    0 & & y(s_i) = y(s_j)\\
    \end{array} \right.
    \label{eq:mi}
\end{equation}
\end{small}

The detailed explanation is listed in Eq.~\ref{eq:mi}. Followed by the definition, our goal is to train a QoE model $\theta$, which enables the relationship of the output $r_\theta$ of the given two sessions, i.e., $r_\theta(o_i)$, $r_\theta(o_j)$, being as consistent as possible with $a_i^j$. In other words, if $a_i^j \neq 0$, we aim to generate a score function $r$ that satisfies:

\begin{small}
\begin{equation}
    a_i^j\left(r_\theta(o_i) - r_\theta(o_j)\right) > 0.
    \label{eq:a}
\end{equation}
\end{small}

\noindent \textbf{Model Architecture.}
As demonstrated in Figure~\ref{fig:qoemodel}, we propose a pairwise learning approach~\cite{ouyang2022training}, denoted as $\theta$. 
Considering the generalization ability, we implement a linear-based and a DNN-based model. 

\begin{figure}
    \centering
    \includegraphics[width=0.85\linewidth]{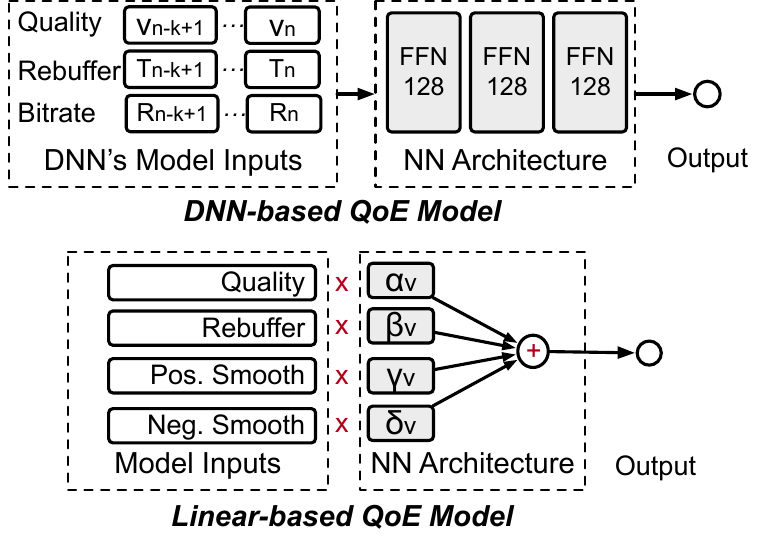}
    \vspace{-10pt}
    \caption{The NN architecture overview for linear-based and DNN-based QoE model. DNN-based model takes a sequential as the input while linear-based model observes the statistics.}
    \vspace{-15pt}
    \label{fig:qoearch}
\end{figure}

\textbf{DNN-based~(QoE$_\texttt{DNN}$):}~As shown in Figure~\ref{fig:qoearch}, our proposed DNN-based QoE model takes $o_n = \{V_n, R_n, T_n\}$ as the inputs, in which $V_n$ are the VMAF~\cite{rassool2017vmaf} of past $k$ chunks, $R_n$ is the sequence of past bitrate picked, and $T_n$ is the past $k$ chunks' rebuffering time. Consistent with the setting of the SQoE-IV, we set $k$ as 7. Furthermore, we take the model's output $r_\theta$ as a single scalar, representing the score for the given video session. The model is simple yet effective, with three feed-forward network (FFN) layers, each with 128 neurons, and a single scalar output without an activation function. We further discuss the performance of applying different feature numbers in \S\ref{sec:abl}.

\textbf{Linear-based~(QoE$_\texttt{lin}$):}~
We choose a linear form that is widely employed in recent studies~\cite{yin2015control}. 
Specifically, the linear-based QoE model, i.e., QoE$_\texttt{lin}$, can be written as: 
\begin{equation}
\label{eq:general-qoe2}
\begin{small}
\begin{aligned}
    {QoE}_\texttt{lin} &=\alpha_v\sum_{n=1}^{^{N}}q(R_{n})-\beta_v\sum_{n=1}^{^{N}}T_{n} -\gamma_v\sum_{n=1}^{^{N-1}}\left[q(R_{n+1})-q(R_{n})\right]_{+} \\
    &-\delta_v\sum_{n=1}^{^{N-1}}\left[q(R_{n+1})-q(R_{n})\right]_{-}.
\end{aligned}
\end{small}
\end{equation}

All the factors are computed as current video quality and rebuffering time, where $\left[q(R_{n+1})-q(R_{n})\right]_{+}$ is positive video quality smoothness, and $\left[q(R_{n+1})-q(R_{n})\right]_{-}$ denotes the negative quality smoothness. 
$\alpha_v$, $\beta_v$, $\gamma_v$, and $\delta_v$ are the learnable parameters. In practice, we consider a ``surrogate scheme'' by first constructing a lightweight NN with a fully connected layer of 4 neurons, corresponding to the weights of the 4 metrics, and outputting a single scalar. Note that we don't apply any activation function here. We then train the NN via vanilla gradient descent. Finally, we obtain the corresponding parameters in the trained network. The linear-based model architecture is also demonstrated in Figure~\ref{fig:qoearch}.

\noindent \textbf{Loss Function.}
Inspired by the training model as DSSM~\cite{huang2013learning}, both DNN-based and linear-based QoE models are optimized via the pairwise-based loss function $\mathcal{L}_{QoE}$, which is listed in Eq.~\ref{sec:loss}. 

\begin{small}
\begin{equation}
    \begin{aligned}
        \mathcal{L}_{QoE} &= \mathbb{E}_{\{o_i, o_j\} \sim q_t, \ q_t \sim Q} [\underbrace{-\log (\sigma(a_i^j(r_\theta(o_i) - r_\theta(o_j))))}_{pairwise \  loss} \\ &+ \underbrace{(1 - |a_i^j|)\left(r_\theta(o_i) - r_\theta(o_j)\right)^2}_{align \  loss} ].
    \end{aligned}
    \label{sec:loss}
\end{equation}
\end{small}

Here $\sigma(\cdot)$ is a Sigmoid function: $\sigma(x) = 1/(1 + e^{-x})$, $r_\theta(\cdot)$ is the output of the proposed QoE model, $a_i^j$ is the relative relationship between the opinion score $y_i$ and $y_j$ (See Eq.~\ref{eq:mi}).
The function consists of two parts. The first part is \emph{pairwise loss}, which is one of the common methods used in the recent RLHF field~\cite{ouyang2022training}. Given two video sessions $o_i$ and $o_j$, this loss function enables the NN to output discriminative results while guaranteeing the same direction as $a_i^j$. The second part is the \emph{align loss}, which aligns the predicted results if $a_i^j=0$. Hence, the \emph{align loss} will be disabled if $a_i^j \neq 0$, and vice versa. Note the network parameters of $r_\theta(o_i)$ and $r_\theta(o_j)$ are shared in our model. 

\noindent \textbf{Training Methodologies.}
Followed by the instruction of Figure~\ref{fig:qoearch}, we summarize the overall training process in Alg.~\ref{alg:qoe}, containing two phases. In detail, the training set is instantly generated during the training process. We randomly pick $K$ samples from different queries, users, and video sessions~(Line 3 - 10). The QoE model is optimized by the generated training set~(Line 11 - 13). We discuss the performance of different batch sizes $K$ in \S\ref{sec:abl}.

\begin{algorithm}
\caption{QoE Model Training Process}
\begin{small}
\begin{algorithmic}[1]
\Require $\theta$: QoE model, $K$: Batch size, $\alpha$: learning rate.
\State Randomly initialize $\theta$.
\While{not done}
\State \setblue{\emph{// Phase 1: Generate training batch.}}
\State Initialize training batch $D$=\{\}.
\For{t in K}
    \State Sample query $q^i \sim Q$, user $u$.
    \State Sample pairs $o_m^i$ and $o_n^i$ from sessions $O^i$ for $q^i$ .
    \State Obtain opinion score $y_u^i(o_m^i)$ and $y_u^i(o_n^i)$ over user $u$.
    \State Compute $a_m^n$ w.r.t $y_u^i(o_m^i)$ and $y_u^i(o_n^i)$~(Eq.~\ref{eq:mi})
    \State $D = D \cup \{o_m^i, o_n^i, a_m^n\}$.
\EndFor
\State \setblue{\emph{// Phase 2: Train QoE model.}}
\For{$\forall \{o_m^i, o_n^i, a_m^n\} \in D$}
    \State \begin{varwidth}[t]{\linewidth}
        Update model using $\mathcal{L}_{QoE}$~(Eq.~\ref{sec:loss}): \par
        \hskip\algorithmicindent $\theta \gets \theta - \alpha \nabla_\theta \mathcal{L}_{QoE}(o_m^i, o_n^i, a_m^n)$.
    \end{varwidth}
\EndFor
\EndWhile
\end{algorithmic}
\end{small}
\label{alg:qoe}
\end{algorithm}

\subsection{QoE Model Validation}
\label{sec:qoemodeleval}
\textbf{Identity Rate.}~After training the QoE model, we observe that recent evaluation metrics such as Spearman’s rank order correlation coefficient~(SRCC) and Pearson’s linear correlation coefficient~(PLCC), which calculate correlations with the MOS, are not suitable for validating our proposed model. Thus, taking inspiration from recent advances in learning reward models~\cite{ouyang2022training}, we introduce a new evaluation metric called the \emph{Identity Rate} to assess the performance of our model.

The metric is motivated by the definition of Eq.~\ref{eq:a}. Given a batch of the testing dataset $D_{test}$, we can compute the Identity Rate as 
\begin{equation}
    \mathbb{E}_{\{i, j\} \sim D_{test}}\mathbb{I}\left(a_i^j(r_\theta(o_i) - r_\theta(o_j)) > 0\right),
\end{equation}
where $\mathbb{I}(\cdot)$ is a binary indicator that returns 1 if the results of the QoE model and opinion scores are in the ``same direction''. In particular, we accept the relative gap between two QoE scores is less than 5\% when $a_i^j=0$~(i.e. a tie).

\begin{table}[]
\centering
\caption{Performance Comparison of QoE Models on SQoE-IV}
\begin{small}
\begin{tabular}{c|ccc}
\toprule
QoE model    & Type   & VQA     & Identity Rate(\%) \\ \hline
Wei et al.~\cite{wei2021reinforcement}   & linear & SSIM~\cite{lin2017structured}       & 60.24              \\
MPC~\cite{yin2015control}   & linear & -       & 61.21              \\ 
Pensieve~\cite{mao2017neural}   & linear & -       & 61.25              \\ 
Puffer~\cite{yan2020learning}       & linear & SSIM$_{db}$~\cite{puffer2022ssimdb} & 62.59              \\
P.1203~\cite{recommendation20171203}       & random forest & - & 64.86              \\
KSQI~\cite{duanmu2019knowledge}       & non-parametric & VMAF~\cite{rassool2017vmaf} & 66.70              \\
Comyco~\cite{huang2019comyco}     & linear & VMAF  & 66.95              \\ 
\rowcolor{mygray}
QoE$_\texttt{lin}$ & linear & VMAF    & 67.26              \\ \hdashline
\emph{MOS-OPT}$^*$  & - & - & \emph{72.15} \\ \hline
\rowcolor{mygray}
QoE$_\texttt{DNN}$    & DNN    & VMAF    & \textbf{75.47}              \\ \bottomrule
\end{tabular}
\end{small}
\label{tbl:model}
\end{table}

\noindent \textbf{Implementation.}~The QoE models are constructed by Tensorflow~\cite{abadi2016tensorflow}. We use Adam~\cite{kingma2014adam} to optimize the NN. As suggested by recent work~\cite{huang2019comyco}, we randomly separate the SQoE-IV database into two parts, 80\% of the database for training~(i.e., 5,771,264 pairs) and 20\% for testing~(i.e., 1,442,941 pairs). 

\begin{figure}
    \centering
    \subfigure[DNN-based model~(QoE$_\texttt{DNN}$)]{
        \includegraphics[width=0.47\linewidth]{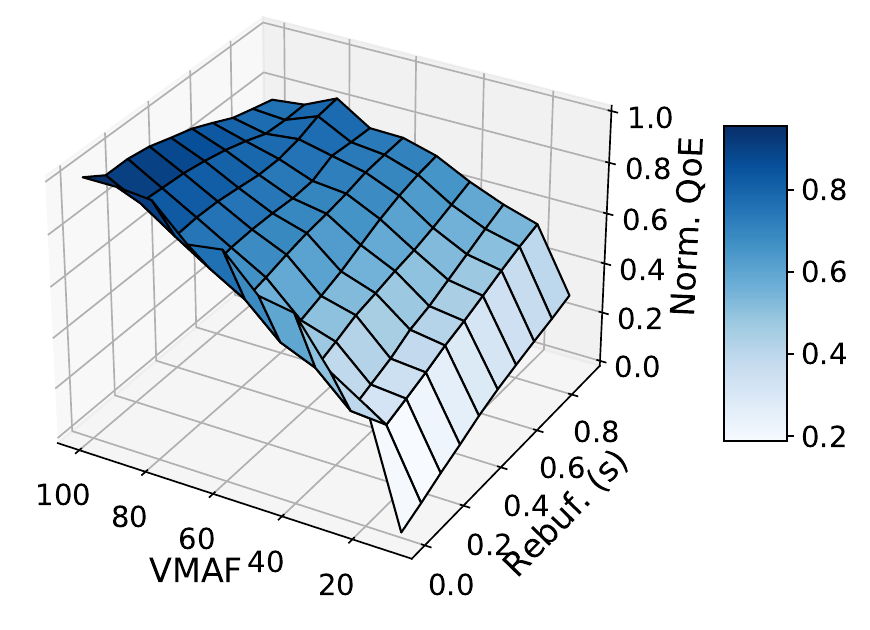}
        \label{fig:dnn3d}
    }
    \subfigure[Linear-based model~(QoE$_\texttt{lin}$)]{
        \includegraphics[width=0.47\linewidth]{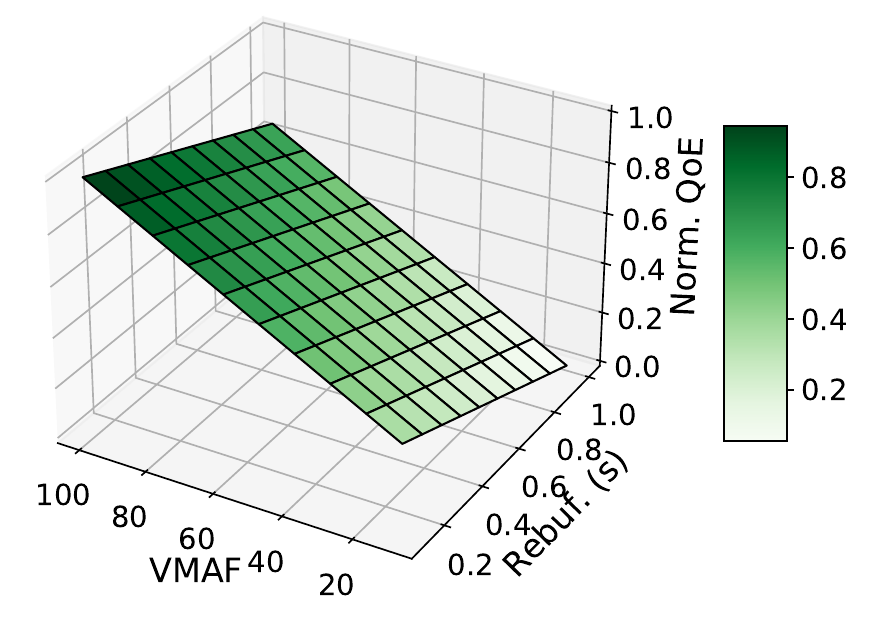}
        \label{fig:lin3d}
    }
    \vspace{-15pt}
    \caption{Visualizing the behaviour of QoE$_\texttt{DNN}$ and QoE$_\texttt{lin}$.}
    \label{fig:vis3d}
\end{figure}

\noindent \textbf{Results.}
The results of the comparison are illustrated in Table~\ref{tbl:model}. 
the QoE models that do not utilize any VQA technique (i.e., bitrate signal only), such as MPC's~\cite{yin2015control} and Pensieve's~\cite{mao2017neural}, exhibit comparable but lower Identity Rates compared with the others. Puffer's QoE model~\cite{yan2020learning} employs SSIM$_{db}$ to outperform Wei et al.'s SSIM-based model~\cite{wei2021reinforcement}. Results of KSQI~\cite{duanmu2019knowledge} and Comyco's~\cite{huang2019comyco} show that VMAF stands for the best VQA metric.

Moreover, we observe that the DNN-based QoE model outperforms all other models, achieving the highest Identity Rate of 75.47\%. The linear-based VMAF model also performs well, with a consistent rate of 67.25\%, making it the best-performing linear model. MOS-OPT, which is offline optimal directly computed by MOS, serves as the upper bound of using MOS. Surprisingly, it even performs worse than the DNN-based QoE model, with a decrease of 4.4\%. Such findings prove that the vanilla MOS-based approach~\cite{huang2019comyco} fails to accurately characterize the users' QoE.

\noindent \textbf{Generalization.}
We visualize the behavior of linear-based and DNN-based models in Figure~\ref{fig:vis3d}. As expected, both two types of models are followed by domain knowledge -- the normalized score is maximum when VMAF is valued at 100 and no rebuffering events. However, we observe that the DNN-based QoE model faces challenges in achieving optimality compared to the linear-based model, with a more fluctuating training process that lacks stability.
For instance, the DNN-based model may predict higher scores even when tethering the VMAF to a fixed value of 10, violating the domain principle of ABR tasks, as rebuffering time increases. 
In contrast, the linear-based model consistently degrades predicted scores with increasing rebuffering time. Similar abnormal improvements are observed with VMAF valued at 80, with noticeable changes occurring at rebuffering times of 0.8.

In summary, the DNN-based model is ``imperfect'', showing unstable performance, while the ``perfect'' linear-based model has acceptable generalization abilities but performs worse than the DNN-based model.

\begin{figure}
    \centering
    \includegraphics[width=0.85\linewidth]{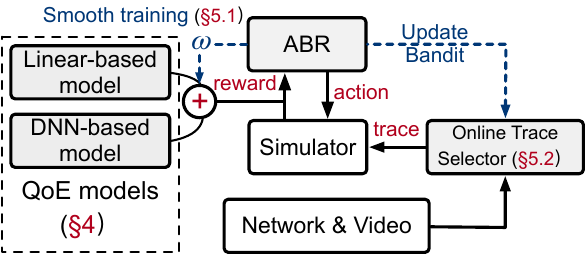}
    \vspace{-10pt}
    \caption{Learned ABR algorithms with the combination of QoE models and online trace selection.}
    \vspace{-15pt}
    \label{fig:jadednn}
\end{figure}

\section{Reinforced ABRs with QoE Models}
\label{sec:rlhf}
In this section, we describe how to generate an NN-based ABR algorithm using trained QoE models.
The overview of the training process is shown in Figure~\ref{fig:jadednn}.
We incorporate two entropy-aware mechanisms: a smooth learning approach and an online trace selection scheme, to fully leverage the advantages of the QoE models.

\subsection{Smooth Training}
\label{sec:smoothtraining}
To avoid the effect of imperfect models, we adopt the linear-based QoE model~(denoted as QoE$_\texttt{lin}$) to train the NN policy in the early stage of training, since the model has a stronger generalization ability which can lead the strategy into the ``near-optimal region''. 
In the later stage, we use the DNN-based QoE model~(denoted as QoE$_\texttt{DNN}$) to continue training the strategy since it can provide more accurate QoE scoring results in such regions. 

Here, an intuitive scheme is to design a two-stage approach that learns the policy with the two QoE models separately in different training phases. However, due to the differences in the meaning of the outputs from the two models, seamless integration of the two models may not be straightforward. 
We, therefore, propose a weighted-decay approach for smooth training. 
The parameter $\omega$ is given to connect the outputs of the linear-based and DNN-based model, thus the surrogate QoE metric can be defined as:
\begin{small}
\begin{equation}
    QoE^{*} = \omega QoE_\texttt{lin} + (1 - \omega) QoE_\texttt{DNN}.
    \label{eq:omega2}
\end{equation}
\end{small}

In this work, $\omega$ is controlled by the entropy of the policy $\phi$ over the trajectory of picked trace $c_t$.
\begin{small}
\begin{equation}
    \omega = \mathbb{E}_{\{s, a\} \sim c_t}\left[\frac{H_{\pi_\phi}(\cdot)}{\log |A|}\right] = \mathbb{E}_{\{s, a\} \sim c_t}\left[\frac{-\mathop{\sum}_{a}\log \pi_\phi(s, a) \pi_\phi(s, a)}{\log |A|}\right].
    \label{eq:omega}
\end{equation}
\end{small}

Here $s$ is the state spaces, $a$ is the bitrate action, and $|A|$ is the number of the bitrate levels. As shown, for any on-policy DRL method~\cite{sutton1998reinforcement}, the entropy of the strategy $H_{\pi_\phi}$ gradually decreases with the training process, as incidentally reduces the weight of QoE$_\texttt{lin}$ and increases the importance of QoE$_\texttt{DNN}$.

\subsection{Online Trace Selection}
\label{sec:OnlineTraceSelection}
Moreover, we use online learning techniques to prioritize a network trace from the training set for training and optimize ABR strategies without prior knowledge. 
To avoid extrapolation errors in the QoE$\texttt{DNN}$ model, we use the policy entropy $H{\pi_{\phi_t}}$ at epoch $t$ as feedback and select the trace with higher entropy outcomes. Higher entropy leads to larger $\omega$ on QoE$^{*}$, which in turn results in a robust and safe QoE metric that is dominated by QoE$_\texttt{lin}$.
In practice, we model the network trace selection problem as a multi-armed bandit problem and propose the module via the discounted UCB (Upper Confidence Bound) algorithm~\cite{garivier2011upper} to pick the network trace with higher entropy outcome for training, with the key method of the module being as follows: 

\begin{small}
\begin{equation}
\label{eq:ct}
    V_t(i) = \frac{\sum_{p=0}^{t} \gamma^{t-p} H_{\pi_{\phi_t}}(\cdot) \mathbb{I}_{\{c_p=i\}} }{\sum_{p=0}^{t} \gamma^{t-p} \mathbb{I}_{\{c_p=i\}}} +  \sqrt{\frac{B\log{t}}{\sum_{p=0}^{t}\mathbb{I}_{\{c_p=i\}}}},
\end{equation}
\end{small}
in which $\gamma$ is the discounted factor, $\mathbb{I}(\cdot)$ will be set as 1 when the trace $i$ has been picked at epoch $p$~(i.e., $c_p=i$), else $\mathbb{I}(\cdot)=0$. Here $B>0$ is a hyper-parameter that controls the probability of exploration. So $V_t(i)$ is defined by the value of network trace $i$ at epoch $t$. For each epoch $t$, the action $c_t$ can be determined as:
\begin{small}
\begin{equation}
    c_t = \mathop{\arg max}_{i} V_t(i).
\end{equation}
\end{small}
In this paper, we set $\gamma=0.999$, $B=0.2$ to balance the trade-off between exploration and exploitation~\cite{zhang2023practical}.

\subsection{Learned Policies with DRL}
We employ a DRL approach to train an ABR policy for maximizing the score obtained from the QoE$^{*}$. Taking Dual-Clip PPO~\cite{ye2019mastering} as the basic training algorithm~(denoted as $\mathcal{L}_\texttt{PPO}$), the combined objective function $\mathcal{L}_{DRL}$ can be expressed as:
\begin{small}
\begin{equation}
    \nabla_\phi \mathcal{L}_\texttt{DRL} = \mathbb{E}_{\{s, a\} \sim c_t}[\nabla_\phi \mathcal{L}_\texttt{PPO}(\phi|s, a, QoE^{*}) + \nabla_\phi \lambda H_{\pi_\phi}(s)],
\end{equation}
\end{small}
where $\lambda$ is an adaptive entropy weight~\cite{li2020suphx} that is dynamically adjusted w.r.t the target entropy $H_{target}$ during training, in which $\alpha$ is the learning rate:
\begin{small}
\begin{equation}
    \lambda = \lambda + \alpha \mathbb{E}_{\{s, a\} \sim c_t}[H_{target}-H_{\pi_\phi}(s)].
\end{equation}
\end{small}

In this paper, \texttt{Jade}'s state $s$ incorporates current buffer occupancy, past chunk's VMAF, past 8 chunks' throughput and download time, and next chunks' video sizes and VMAF scores. The action $a$ is a vector indicating the bitrate selection probabilities.
To capture features from diverse input types, \texttt{Jade}'s actor-critic network~\cite{sutton2018reinforcement} adopts a combination of multiple FFN layers with 128 neurons to extract and combine the underlying features. 
\texttt{Jade}'s learning tools are built with TensorFlow 2.8.1~\cite{abadi2016tensorflow}. We set learning rate $\alpha=10^{-4}$, $H_{target}=0.1$.


\begin{figure*}
    \centering
    \subfigure[VMAF vs. Stall Ratio]{
        \includegraphics[width=0.23\linewidth]{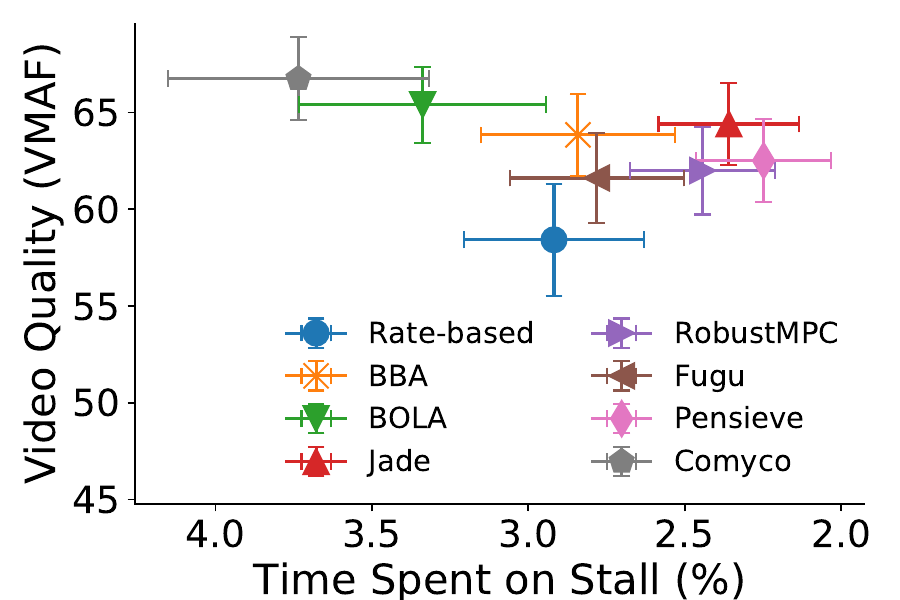}
        \label{fig:norwaybr}
    }
    \subfigure[VMAF vs. VMAF Change]{
        \includegraphics[width=0.23\linewidth]{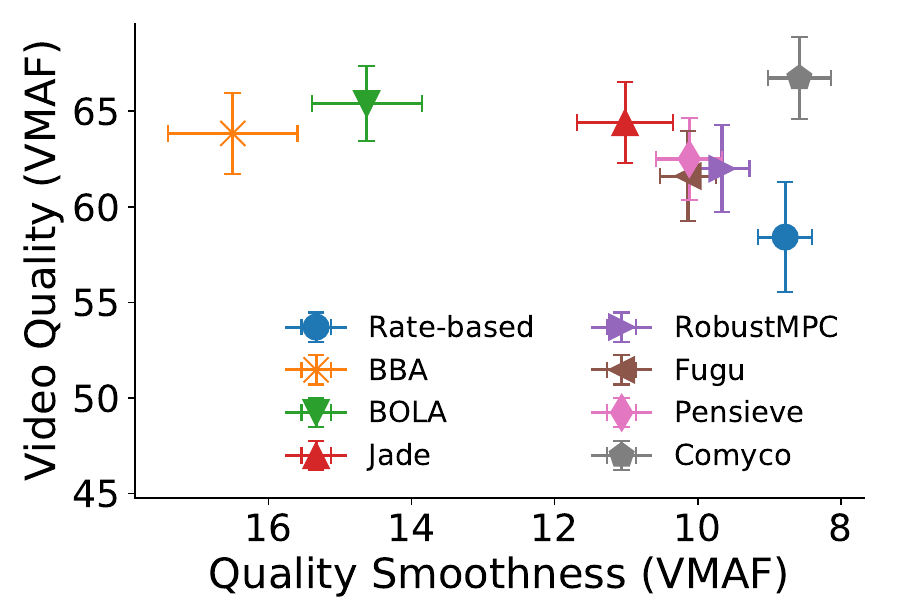}
        \label{fig:norwaybs}
    }
    \subfigure[QoE$_\texttt{DNN}$ vs. Buffer]{
        \includegraphics[width=0.23\linewidth]{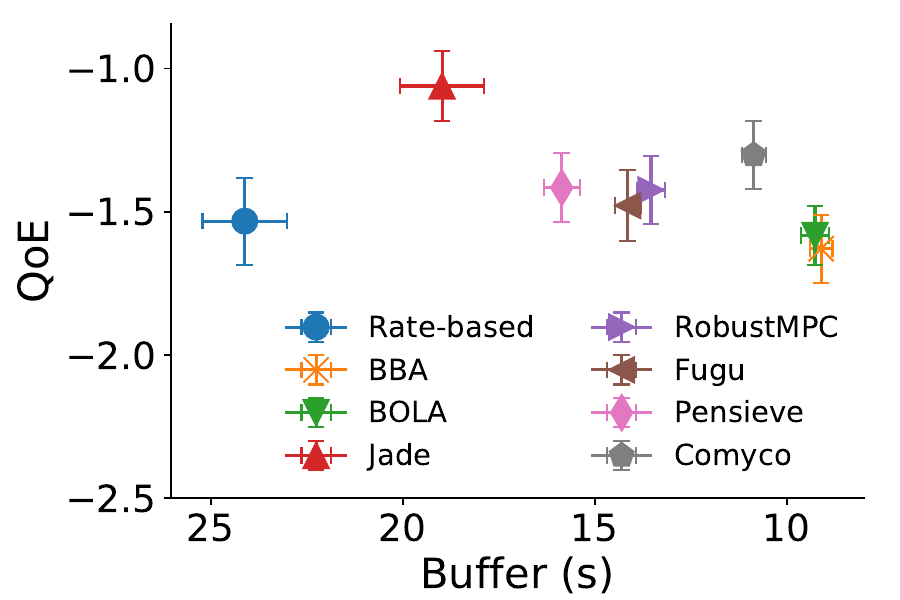}
        \label{fig:norwaysr}
    }
    \subfigure[CDF of QoE$_\texttt{DNN}$]{
        \includegraphics[width=0.23\linewidth]{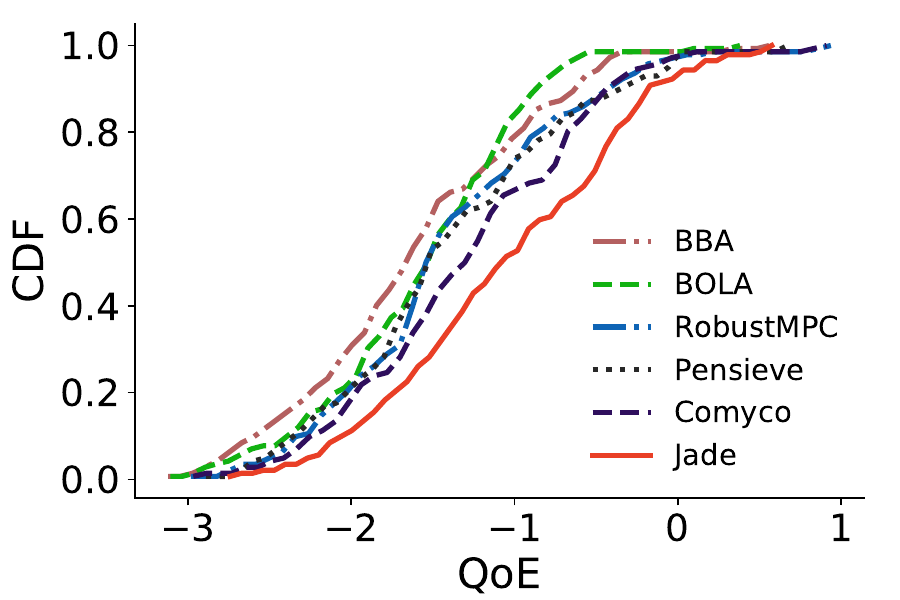}
        \label{fig:norwayqoe}
    }
    \vspace{-15pt}
    \caption{Comparing \texttt{Jade} with recent ABR algorithms over the HSDPA dataset. Error bars show 95\% confidence intervals.}
    \vspace{-5pt}
    \label{fig:norway}
\end{figure*}

\begin{figure*}
    \centering
    \subfigure[VMAF vs. Stall Ratio]{
        \includegraphics[width=0.23\linewidth]{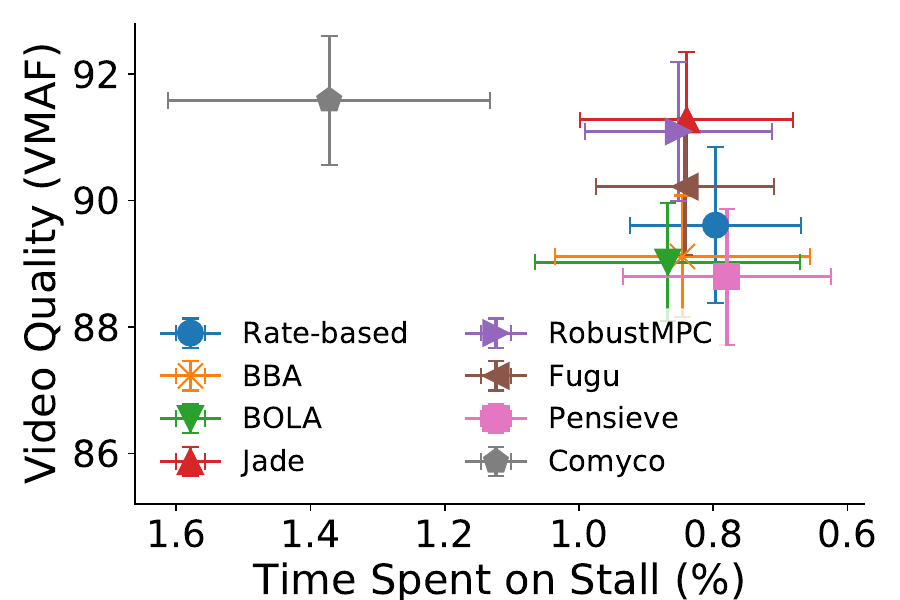}
        \label{fig:fccbr}
    }
    \subfigure[VMAF vs. VMAF Change]{
        \includegraphics[width=0.23\linewidth]{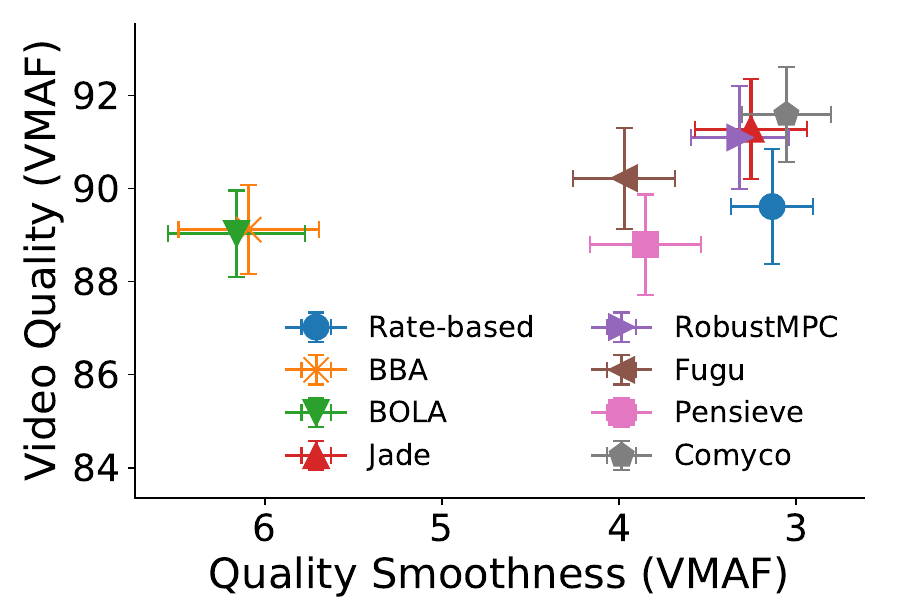}
        \label{fig:fccbs}
    }
    \subfigure[QoE$_\texttt{DNN}$ vs. Buffer]{
        \includegraphics[width=0.23\linewidth]{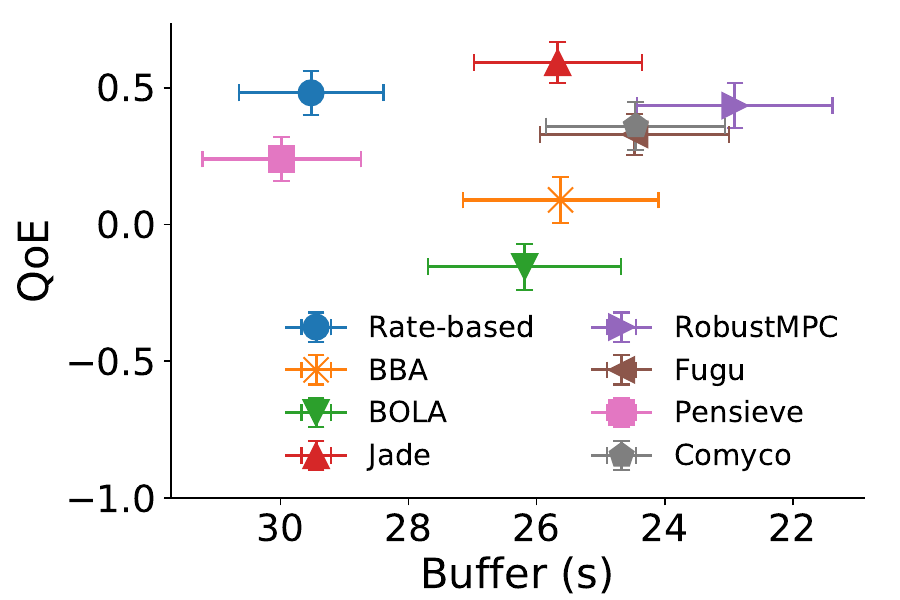}
        \label{fig:fccsr}
    }
    \subfigure[CDF of QoE$_\texttt{DNN}$]{
        \includegraphics[width=0.23\linewidth]{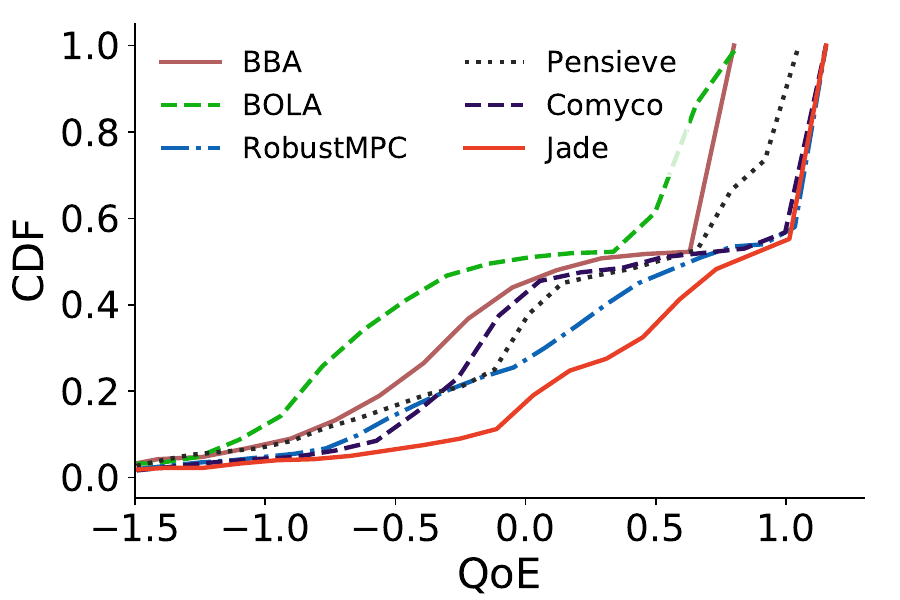}
        \label{fig:fccqoe}
    }
    \vspace{-15pt}
    \caption{Comparing \texttt{Jade} with existing ABR algorithms using QoE$_\texttt{DNN}$. Results are collected over the FCC-18 dataset.}
    \vspace{-10pt}
    \label{fig:fcc}
\end{figure*}

\section{Evaluation}
\label{sec:eval}
\subsection{Methodology}
\label{sec:Implementation}
\noindent \textbf{Experimental Setup.}~We employ trace-driven simulation with virtual player~\cite{mao2019park}, adopt Comyco video dataset~\cite{comycovideogithub} for training and EnvivioDash3~\cite{envivio2016} for testing. All videos are encoded using H.264~\cite{wiegand2003overview} with bitrates of \{0.3, 0.75, 1.2, 1.85, 2.85, 4.3\} Mbps.
Note that \texttt{Jade} supports any other simulators, e.g., CasualSim~\cite{alomar2023causalsim}.

\noindent \textbf{Network Trace Datasets.} 
\label{sec:NetworkDatasets}
We utilize multiple trace datasets, including Pensieve training set~\cite{mao2017neural}, Puffer public dataset~\cite{yan2020learning}, and the 5G dataset~\cite{Narayanan2020lumos5g} to train \texttt{Jade}. The dataset contains a large corpus of network traces, covering 0.1 to 100 Mbps. 
For testing, we employ two types of public datasets: slow-network paths~($\leq$6Mbps)~(HSDPA~\cite{riiser2013commute} and FCC~\cite{bworld}) and fast-network paths~(FCC-18~\cite{meng2021practically}).

\noindent \textbf{QoE Metrics.}
\label{sec:QoE}
We employ two QoE metrics. The first is QoE$_\texttt{lin}$~\cite{yin2015control, akhtar2018oboe, mao2017neural,yan2020learning}, the linear-based QoE model.
The QoE parameters are set as $\alpha_v=0.535$, $\beta_v=-0.215$, $\gamma_v=0.13$, $\delta_v=1.37$.
The second is QoE$_\texttt{DNN}$, the trained DNN-based model mentioned in \S\ref{sec:pairwise}.
Note all the results of the paper are summarized as QoE$_\texttt{DNN}$.

\noindent \textbf{ABR Baselines.}
\label{sec:abrbaseline}
We select representational ABR algorithms from various fundamental principles. 
For QoS-driven ABR algorithms, that include \textbf{BBA}~\cite{huang2015buffer} and \textbf{BOLA}~\cite{spiteri2020bola}:~the traditional Buffer-based approach, \textbf{Rate-based}~\cite{jiang2014improving}: a throughput-based scheme via harmonic mean.
For QoE-driven ABRs, we choose:~\textbf{RobustMPC}~\cite{yin2015control}: make bitrate decisions by solving a QoE optimization problem for the next 5 chunks; \textbf{Pensieve}~\cite{mao2017neural}:~the DRL-based ABR scheme to learn
bitrate decisions for maximizing a QoE reward; \textbf{Comyco}~\cite{huang2019comyco}:~a learning-based ABR scheme to maximize QoE objectives via imitation learning; 
\textbf{Fugu}~\cite{yan2020learning}:~a hybrid ABR algorithm using DNN for prediction and MPC for decision-making.
In addition, we also integrate our proposed QoE models into recent schemes, including \textbf{MPC-lin} and \textbf{MPC-DNN}: perform RobustMPC via maximizing QoE$_\texttt{lin}$ and QoE$_\texttt{DNN}$; \textbf{Comyco-lin}: mimic policies with QoE$_\texttt{lin}$; \textbf{Jade-lin} and \textbf{Jade-DNN}: reinforced ABR algorithms towards better QoE$_\texttt{lin}$ and QoE$_\texttt{DNN}$ reward; \textbf{PPO-ptx}~\cite{ouyang2022training}: a fine-tuning approach towards better QoE$_\texttt{DNN}$~(i.e., the default scheme for ChatGPT~\cite{chatgpt2023}).

\subsection{\texttt{Jade} vs. Existing ABR algorithms}
\label{sec:slowfast}
We train \texttt{Jade} once and employ trace-driven simulation to conduct a performance comparison between \texttt{Jade} and several existing ABR algorithms across slow-network and fast-network paths.

\noindent \textbf{Slow-network paths.}~Figure~\ref{fig:norway} shows the overall performance of \texttt{Jade} and recent works over the HSDPA network traces. We see that \texttt{Jade} stands for the best approach, as its QoE outperforms existing ABR algorithms by 22.56\%-38.13\%~(see Figure~\ref{fig:norwayqoe}). Specifically, it performs better than state-of-the-art heuristic RobustMPC~(29.22\%) and RL-based approach Pensieve~(28.82\%) -- they are all optimized by the linear combination of QoS metrics~\cite{yin2015control}, resulting in similar performances as well. Comparing the performance with the closest scheme Comyco, we find that despite being 22.56\% lower than \texttt{Jade}, it has relatively improved by 8.09\% compared to other algorithms. 
By analyzing the Identity Rate of each QoE model as shown in Table~\ref{tbl:model}, we highlight the importance of using an accurate QoE model as the reward, as it is positively correlated with the performance of the learned ABR algorithm.

Furthermore, Figure~\ref{fig:norwaybr} demonstrates that almost all the ABR schemes are performed within the acceptable region,  with a stall ratio of less than 5\%~\cite{narayanan2021variegated}. Among them, Comyco, BOLA, \texttt{Jade}, and Pensieve achieve the Pareto Frontier, indicating that they provide optimal policies in terms of trade-offs between different objectives, while \texttt{Jade} reaches the best QoE performance, standing for the Top-2 scheme in terms of QoS as well. By referring to Figure~\ref{fig:norwaybs} and Figure~\ref{fig:norwaysr}, we can see the detailed behavior in terms of buffer occupancy and quality smoothness. While \texttt{Jade} may not reach the upper right corner of the region, it consistently makes bitrate decisions within a reasonable range for QoS considerations, since it ranks at mid-level among all candidates.

\noindent \textbf{Fast-network paths.}~At the same time, Figure~\ref{fig:fcc} illustrates the performance of \texttt{Jade} and the baselines on the FCC-18 network dataset, which represents a wide range of network conditions that closely resemble real-world networks. Unsurprisingly, the CDF curve of QoE, as depicted in Figure~\ref{fig:fccqoe}, shows that \texttt{Jade} surpasses other ABR schemes, achieving significant improvements in QoE by at least 23.07\%. Impressively, \texttt{Jade} improves the average QoE$_\texttt{DNN}$ by 36.32\%, 64.66\%, 79.95\%, and 1.47$\times$ compared to state-of-the-art QoE-driven ABR algorithms such as RobustMPC, Comyco, Fugu, and Pensieve, respectively.
With further investigation on QoS in Figure~\ref{fig:fccbr}, we observe that \texttt{Jade} can well balance the trade-off between quality and stall ratio, maintaining the session within an acceptable buffer range~(see Figure~\ref{fig:fccqoe}). Moreover, \texttt{Jade} also shows outstanding QoE improvements in the fast-network paths, ranging from 1.86\% to 2.53\%, compared to QoS-driven ABR algorithms, such as Rate-based, BBA, and BOLA. 
At the same time, Figure~\ref{fig:fccbs} reveals that \texttt{Jade} slightly performs better than the closest QoE-driven approach RobustMPC~(+0.2\% quality, -1.14\% stall ratio, -1.9\% smoothness) in terms of QoS metrics.
However, it significantly outperforms RobustMPC in terms of QoE, which indicates that the QoS result does not necessarily represent QoE -- the duration of buffering and bitrate switching events directly impact users' experience, which cannot be reflected by traditional linear-based QoE models like QoE$_\texttt{lin}$.
Thus, applying a sequential QoE model like QoE$_\texttt{DNN}$ is reasonable, which can not only consider the ``thermodynamics'' of the outcome but also appreciate the ``kinetics'' of the process. 

\begin{figure}
    \centering
    \includegraphics[width=0.48\linewidth]{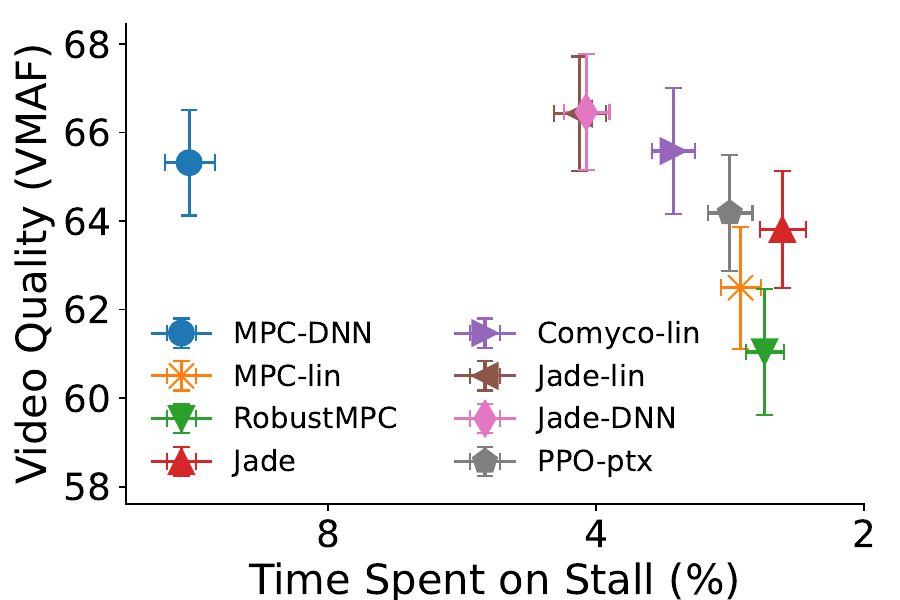}
    \includegraphics[width=0.48\linewidth]{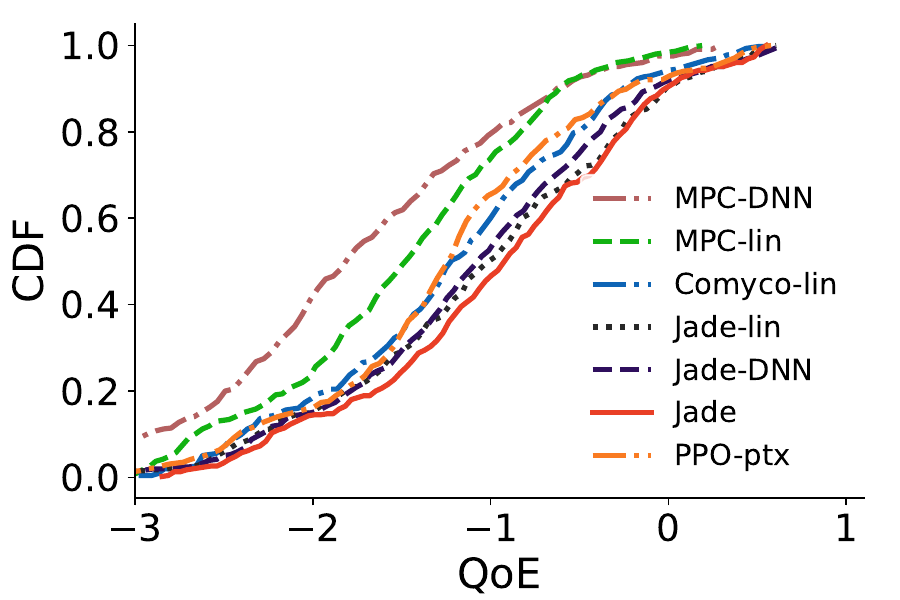}
    \vspace{-10pt}
    \caption{Performance comparison of \texttt{Jade} and several QoE-driven ABRs. Results are collected over the FCC dataset.}
    \vspace{-15pt}
    \label{fig:jadeqoedriven}
\end{figure}

\subsection{\texttt{Jade} vs. other QoE-driven ABR algorithms}
\label{sec:jadevsqoe}
We compare the performance of \texttt{Jade} with existing QoE-driven ABR approaches in Figure~\ref{fig:jadeqoedriven}, where the results are collected from the FCC network dataset. Note MPC is RobustMPC~\cite{yin2015control}.
Here we show three key findings. 
First, after changing the goal of MPC to maximize QoE$_\texttt{DNN}$~(i.e., MPC-DNN), the overall performance drops significantly, almost doubling the stall ratio compared with MPC but optimized by QoE$_\texttt{lin}$~(i.e., MPC-lin). Such results indicate that we should exercise greater caution in using such ``imperfect QoE models''. Otherwise, unwavering trust in these models may result in misguided strategies, leading to undesired outcomes. As such, the utilization of QoE$_\texttt{DNN}$ to enhance heuristics remains an open question. 
Secondly, both Comyco-lin and Jade-lin employ QoE$_\texttt{lin}$ as the reward signal, leading to a similar performance in regard to video quality and stall ratio -- the only difference between those two is the training methodologies.
Thirdly, \texttt{Jade} improves QoE$_\texttt{DNN}$ by 18.07\% compared with PPO-ptx, while experiencing a decrease of 12.72\% in stall ratio, demonstrating the strengths of learning from \emph{clean slate}.
Furthermore, Jade-DNN performs worse than Jade-lin, with the decreases on average QoE$_\texttt{DNN}$ of 21.17\%. The key reason is Jade-DNN occurs too many rebuffering events~(stall ratio: 4.1\%) and fails to obtain higher video quality, just like MPC-DNN~(stall ratio: 11.47\%). 
Finally, through the fusion of linear-based and DNN-based QoE models, \texttt{Jade} not only attains the Pareto Frontier on QoS but also achieves superior QoE outcomes across all considered scenarios, demonstrating the superiority against recent schemes.

\subsection{Ablation Study}
\label{sec:abl}
\newcolumntype{g}{>{\columncolor{mygray}}c}
\begin{table}[]
\caption{Sweeping the parameters for QoE$_\texttt{DNN}$.}
\vspace{-10pt}
\label{tbl:gpt}
\begin{small}
\begin{tabular}{c|c|ccccg}
\toprule
   \multirow{2}{*}{\begin{tabular}[c]{@{}c@{}}Feat. Num. \\ Per Layer\end{tabular}} & \multirow{2}{*}{\begin{tabular}[c]{@{}c@{}}InstructGPT \\ \cite{ouyang2022training}\end{tabular}} & \multicolumn{5}{c}{\emph{Batch Size K}}        \\ \cline{3-7} 
   &                          & 512   & 1024  & 2048  & 4096  & 8192  \\ \hline
64 & 73.31                    & 74.68 & 74.66 & 74.86 & 74.99 & 75.11 \\
\rowcolor{mygray}
128 & 73.54                   & 75.42 & 75.45 & 75.46 & 75.46 & \textbf{75.47} \\
256 & 73.60                   & 75.44 & 75.44 & 75.42 & 75.45 & 75.46 \\ \bottomrule
\end{tabular}
\end{small}
\end{table}

We conduct several experiments to better understand \texttt{Jade}'s hyper-parameter settings and entropy-aware mechanisms~(\S\ref{sec:rlhf}).

\noindent \textbf{Comparison for QoE models.}
Table~\ref{tbl:gpt} shows that QoE$_\texttt{DNN}$ consistently outperforms the training methodologies from InstructGPT~\cite{ouyang2022training}, achieving the highest Identity Rate~(\S\ref{sec:qoemodeleval}) of 75.47\% with a batch size of 8192 and a feature number of 128. Additionally, setting the feature number to 256 rivals the 128 version but incurs higher computational costs. Thus, optimal parameter settings for the model include larger batch sizes and a feature number of 128.

\begin{figure}
    \centering
    \includegraphics[width=0.48\linewidth]{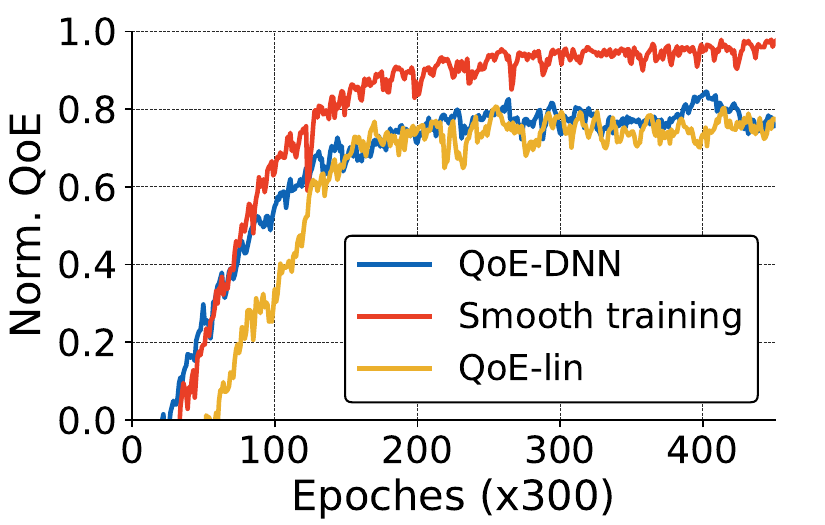}
    \includegraphics[width=0.48\linewidth]{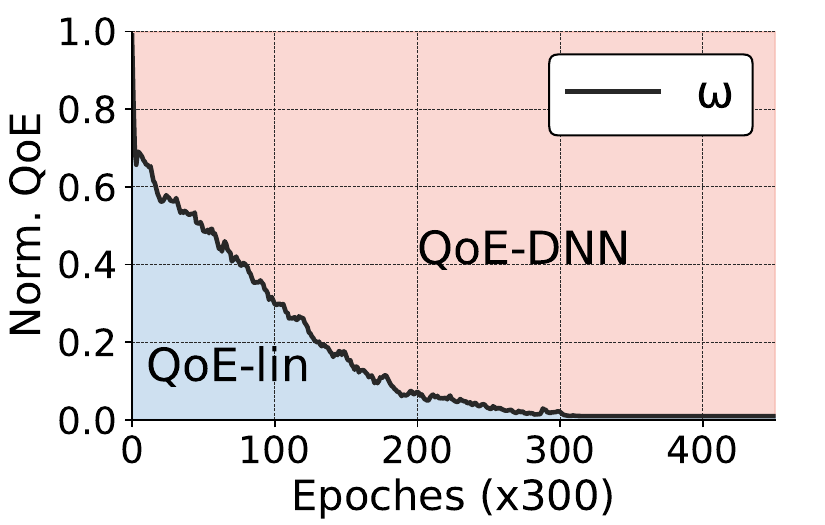}
    \vspace{-10pt}
    \caption{The learning curve of smooth training.}
    \vspace{-10pt}
    \label{fig:ent2}
\end{figure}

\begin{figure}
    \centering
    \includegraphics[width=0.48\linewidth]{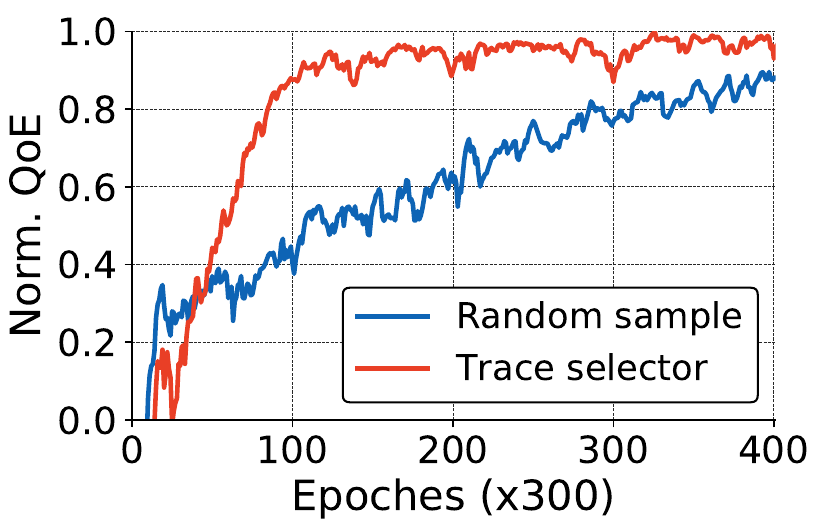}
    \includegraphics[width=0.48\linewidth]{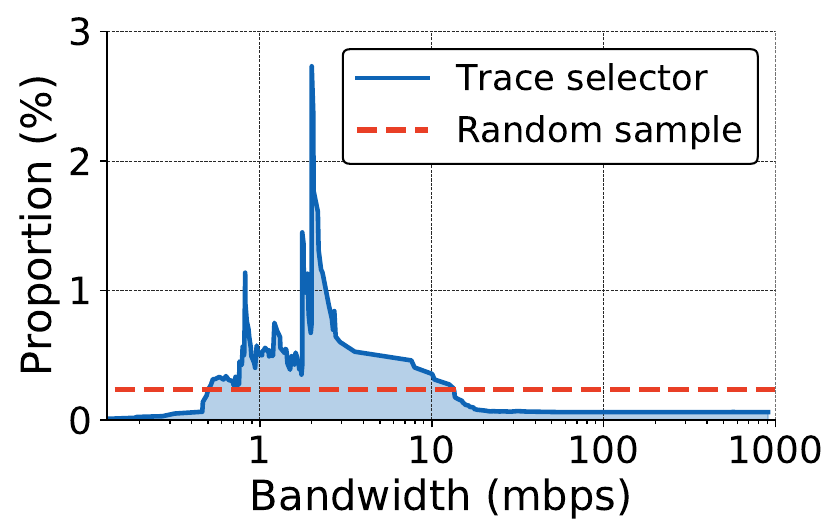}
    \vspace{-10pt}
    \caption{Visualizing online trace selection.}
    \vspace{-10pt}
    \label{fig:prop2}
\end{figure}

\noindent \textbf{Validation for smooth training.}
Figure~\ref{fig:ent2} reports the normalized learning curves of applying different QoE models, in which \emph{smooth training} follows the principle described in \S\ref{sec:smoothtraining}. The results are reported over the validation set every 300 epochs.
As demonstrated, leveraging smooth training enables the learned policy to achieve better performances compared with using QoE$_\texttt{DNN}$ and QoE$_\texttt{lin}$ solely~(25\% on QoE). Moreover, the evolution curve of $\omega$ indicates that QoE$_\texttt{DNN}$ dominates the majority of the training process, while QoE$_\texttt{lin}$ also plays an important role at the beginning, covering approximately 75\% of the training process.
Such results further demonstrate that both QoE models are indispensable.

\noindent \textbf{Effectiveness of trace selector.}
We present the detailed results of utilizing and not utilizing (i.e., random sample) the trace selector in Figure~\ref{fig:prop2}. Note that we also summarize the performance over the validation set every 300 epochs.
Results show that selecting useful traces not only improves overall performance but also accelerates the learning process. Moreover, The PDF curve of the trace selection demonstrates varying probabilities of selecting different traces during training. Traces with an average bandwidth of 1-3 Mbps, technically covering the bitrate ladders~(\S\ref{sec:Implementation}), are frequently chosen, which shows the effectiveness of the proposed selector.

\vspace{-10pt}
\section{Related work}
\label{sec:related}
\noindent \textbf{QoE Metrics.}~Recent QoE models prove the effectiveness of machine learning techniques in mapping \emph{streaming features} to \emph{mean opinion scores}, such as regressive models~\cite{huang2019comyco,turkkan2022greenabr}, DNNs~\cite{du2020video}, and random forests~\cite{recommendation20171203}. 
However, such schemes may not accurately reflect users' opinions due to the existence of user heterogeneity, while rank-based QoE models diverge~(\S\ref{sec:qoemodeleval}).

\noindent \textbf{QoS-driven ABR approaches} primarily employ observed metrics from different perspectives to achieve better QoS performances~\cite{huang2019hindsight}. 
Throughput-based ABR algorithms, such as FESTIVE~\cite{jiang2014improving} and PANDA~\cite{li2014probe}, adopt past throughput to forecast future bandwidth and choose the appropriate bitrate. 
BBA~\cite{huang2015buffer}, QUETRA~\cite{yadav2017quetra}, and BOLA~\cite{spiteri2020bola} are buffer-based approaches that primarily use the buffer occupancy to select the bitrate. 
Zwei~\cite{huang2020self,huang2021zwei} uses self-play learning methods to directly meet QoS requirements.
However, all of these approaches optimize QoS rather than QoE.

\noindent \textbf{QoE-driven ABR algorithms} adjust the bitrate based on observed network status to fulfill users' QoE.
MPC~\cite{yin2015control} employs control theory to maximize QoE via harmonic throughput prediction, while Fugu~\cite{yan2020learning} trains the DNN-based throughput predictor periodically. Pensieve~\cite{mao2017neural} and Comyco~\cite{huang2019comyco,huang2020quality} use DRL and imitation learning to generate DNNs without prior knowledge. MERINA~\cite{kan2022improving} and A$^2$BR~\cite{huang2022learning} quickly adapt to different networks. However, the aforementioned works are optimized by linear-based QoE models, which is not suitable for DNN-based models~(like QoE$_\texttt{DNN}$, see \S\ref{sec:rlhf}). 

In addition, recent works also focus on the diversity of QoE~\cite{zhang2020leveraging, qiao2020beyond, zhang2021sensei,zuo2022adaptive}. While \texttt{Jade} provides a satisfactory ABR algorithm for ``average users'' based on rank-based QoE models~(\S\ref{sec:pairwise}).

\vspace{-5pt}
\section{Conclusion}
Off-the-shelf QoE-driven ABR algorithms, optimized by MOS-based QoE models, may not account for user heterogeneity. We developed a novel ABR approach called \texttt{Jade}, which used the main concept of RLHF with a rank-based QoE model and DRL with a hybrid feedback mechanism. In detail, \texttt{Jade} addressed challenges in training two rank-based QoE models with different architectures and generating entropy-aware ABR algorithms for smooth training. Evaluations showed that \texttt{Jade} outperforms existing algorithms in both slow- and fast-network paths and achieves the Pareto Frontier. Further research will focus on treating user engagement as the opinion score~\cite{zhang2022enabling} and applying \texttt{Jade} into the live streaming scenarios~\cite{yan2020learning}.

\noindent \textbf{Acknowledgements.}~We thank the reviewers for the valuable feedback. To Comyco, \texttt{Jade} is her favorite accessory. This work was supported by NSFC under Grant 61936011.

\bibliographystyle{ACM-Reference-Format}
\balance
\bibliography{ref}


\begin{thebibliography}{76}


\ifx \showCODEN    \undefined \def \showCODEN     #1{\unskip}     \fi
\ifx \showDOI      \undefined \def \showDOI       #1{#1}\fi
\ifx \showISBNx    \undefined \def \showISBNx     #1{\unskip}     \fi
\ifx \showISBNxiii \undefined \def \showISBNxiii  #1{\unskip}     \fi
\ifx \showISSN     \undefined \def \showISSN      #1{\unskip}     \fi
\ifx \showLCCN     \undefined \def \showLCCN      #1{\unskip}     \fi
\ifx \shownote     \undefined \def \shownote      #1{#1}          \fi
\ifx \showarticletitle \undefined \def \showarticletitle #1{#1}   \fi
\ifx \showURL      \undefined \def \showURL       {\relax}        \fi
\providecommand\bibfield[2]{#2}
\providecommand\bibinfo[2]{#2}
\providecommand\natexlab[1]{#1}
\providecommand\showeprint[2][]{arXiv:#2}

\bibitem[\protect\citeauthoryear{??}{env}{2016}]%
        {envivio2016}
 \bibinfo{year}{2016}\natexlab{}.
\newblock \bibinfo{title}{EnvivioDash3}.
\newblock
  \bibinfo{howpublished}{https://dash.akamaized.net/envivio/EnvivioDash3/}.
\newblock


\bibitem[\protect\citeauthoryear{??}{HLS}{2019}]%
        {HLS}
 \bibinfo{year}{2019}\natexlab{}.
\newblock \bibinfo{title}{HTTP Live Streaming}.
\newblock \bibinfo{howpublished}{\url{https://developer.apple.com/streaming/}}.
\newblock


\bibitem[\protect\citeauthoryear{Abadi, Barham, Chen, Chen, Davis, Dean, Devin,
  Ghemawat, Irving, Isard, et~al\mbox{.}}{Abadi et~al\mbox{.}}{2016}]%
        {abadi2016tensorflow}
\bibfield{author}{\bibinfo{person}{Mart{\'\i}n Abadi}, \bibinfo{person}{Paul
  Barham}, \bibinfo{person}{Jianmin Chen}, \bibinfo{person}{Zhifeng Chen},
  \bibinfo{person}{Andy Davis}, \bibinfo{person}{Jeffrey Dean},
  \bibinfo{person}{Matthieu Devin}, \bibinfo{person}{Sanjay Ghemawat},
  \bibinfo{person}{Geoffrey Irving}, \bibinfo{person}{Michael Isard},
  {et~al\mbox{.}}} \bibinfo{year}{2016}\natexlab{}.
\newblock \showarticletitle{TensorFlow: A System for Large-Scale Machine
  Learning.}. In \bibinfo{booktitle}{\emph{OSDI}}, Vol.~\bibinfo{volume}{16}.
  \bibinfo{pages}{265--283}.
\newblock


\bibitem[\protect\citeauthoryear{Adams and Marshall}{Adams and
  Marshall}{1996}]%
        {adams1996developmental}
\bibfield{author}{\bibinfo{person}{Gerald~R Adams} {and}
  \bibinfo{person}{Sheila~K Marshall}.} \bibinfo{year}{1996}\natexlab{}.
\newblock \showarticletitle{A developmental social psychology of identity:
  Understanding the person-in-context}.
\newblock \bibinfo{journal}{\emph{Journal of adolescence}}
  \bibinfo{volume}{19}, \bibinfo{number}{5} (\bibinfo{year}{1996}),
  \bibinfo{pages}{429--442}.
\newblock


\bibitem[\protect\citeauthoryear{Akhtar, Nam, Govindan, et~al\mbox{.}}{Akhtar
  et~al\mbox{.}}{2018}]%
        {akhtar2018oboe}
\bibfield{author}{\bibinfo{person}{Zahaib Akhtar}, \bibinfo{person}{Yun~Seong
  Nam}, \bibinfo{person}{Ramesh Govindan}, {et~al\mbox{.}}}
  \bibinfo{year}{2018}\natexlab{}.
\newblock \showarticletitle{Oboe: auto-tuning video ABR algorithms to network
  conditions}. In \bibinfo{booktitle}{\emph{SIGCOMM 2018}}. ACM,
  \bibinfo{pages}{44--58}.
\newblock


\bibitem[\protect\citeauthoryear{Alomar, Hamadanian, Nasr-Esfahany, Agarwal,
  Alizadeh, and Shah}{Alomar et~al\mbox{.}}{2023}]%
        {alomar2023causalsim}
\bibfield{author}{\bibinfo{person}{Abdullah Alomar}, \bibinfo{person}{Pouya
  Hamadanian}, \bibinfo{person}{Arash Nasr-Esfahany}, \bibinfo{person}{Anish
  Agarwal}, \bibinfo{person}{Mohammad Alizadeh}, {and}
  \bibinfo{person}{Devavrat Shah}.} \bibinfo{year}{2023}\natexlab{}.
\newblock \showarticletitle{$\{$CausalSim$\}$: A Causal Framework for Unbiased
  $\{$Trace-Driven$\}$ Simulation}. In \bibinfo{booktitle}{\emph{20th USENIX
  Symposium on Networked Systems Design and Implementation (NSDI 23)}}.
  \bibinfo{pages}{1115--1147}.
\newblock


\bibitem[\protect\citeauthoryear{Bai, Jones, Ndousse, Askell, Chen, DasSarma,
  Drain, Fort, Ganguli, Henighan, et~al\mbox{.}}{Bai et~al\mbox{.}}{2022}]%
        {bai2022training}
\bibfield{author}{\bibinfo{person}{Yuntao Bai}, \bibinfo{person}{Andy Jones},
  \bibinfo{person}{Kamal Ndousse}, \bibinfo{person}{Amanda Askell},
  \bibinfo{person}{Anna Chen}, \bibinfo{person}{Nova DasSarma},
  \bibinfo{person}{Dawn Drain}, \bibinfo{person}{Stanislav Fort},
  \bibinfo{person}{Deep Ganguli}, \bibinfo{person}{Tom Henighan},
  {et~al\mbox{.}}} \bibinfo{year}{2022}\natexlab{}.
\newblock \showarticletitle{Training a helpful and harmless assistant with
  reinforcement learning from human feedback}.
\newblock \bibinfo{journal}{\emph{arXiv preprint arXiv:2204.05862}}
  (\bibinfo{year}{2022}).
\newblock


\bibitem[\protect\citeauthoryear{Bampis, Li, Katsavounidis, Huang, Ekanadham,
  and Bovik}{Bampis et~al\mbox{.}}{2018}]%
        {bampis2018towards}
\bibfield{author}{\bibinfo{person}{Christos~G Bampis}, \bibinfo{person}{Zhi
  Li}, \bibinfo{person}{Ioannis Katsavounidis}, \bibinfo{person}{Te-Yuan
  Huang}, \bibinfo{person}{Chaitanya Ekanadham}, {and} \bibinfo{person}{Alan~C
  Bovik}.} \bibinfo{year}{2018}\natexlab{}.
\newblock \showarticletitle{Towards perceptually optimized end-to-end adaptive
  video streaming}.
\newblock \bibinfo{journal}{\emph{arXiv preprint arXiv:1808.03898}}
  (\bibinfo{year}{2018}).
\newblock


\bibitem[\protect\citeauthoryear{Bentaleb, Begen, and Zimmermann}{Bentaleb
  et~al\mbox{.}}{2016}]%
        {bentaleb2016sdndash}
\bibfield{author}{\bibinfo{person}{Abdelhak Bentaleb}, \bibinfo{person}{Ali~C
  Begen}, {and} \bibinfo{person}{Roger Zimmermann}.}
  \bibinfo{year}{2016}\natexlab{}.
\newblock \showarticletitle{SDNDASH: Improving QoE of HTTP adaptive streaming
  using software defined networking}. In \bibinfo{booktitle}{\emph{Proceedings
  of the 24th ACM international conference on Multimedia}}.
  \bibinfo{pages}{1296--1305}.
\newblock


\bibitem[\protect\citeauthoryear{Bentaleb, Taani, Begen, Timmerer, and
  Zimmermann}{Bentaleb et~al\mbox{.}}{2018}]%
        {bentaleb2018survey}
\bibfield{author}{\bibinfo{person}{Abdelhak Bentaleb}, \bibinfo{person}{Bayan
  Taani}, \bibinfo{person}{Ali~C Begen}, \bibinfo{person}{Christian Timmerer},
  {and} \bibinfo{person}{Roger Zimmermann}.} \bibinfo{year}{2018}\natexlab{}.
\newblock \showarticletitle{A Survey on Bitrate Adaptation Schemes for
  Streaming Media over HTTP}.
\newblock \bibinfo{journal}{\emph{IEEE Communications Surveys \& Tutorials}}
  (\bibinfo{year}{2018}).
\newblock


\bibitem[\protect\citeauthoryear{Chen, Maguluri, Shakkottai, and
  Shanmugam}{Chen et~al\mbox{.}}{2021}]%
        {chen2021lyapunov}
\bibfield{author}{\bibinfo{person}{Zaiwei Chen}, \bibinfo{person}{Siva~Theja
  Maguluri}, \bibinfo{person}{Sanjay Shakkottai}, {and}
  \bibinfo{person}{Karthikeyan Shanmugam}.} \bibinfo{year}{2021}\natexlab{}.
\newblock \showarticletitle{A Lyapunov theory for finite-sample guarantees of
  asynchronous Q-learning and TD-learning variants}.
\newblock \bibinfo{journal}{\emph{arXiv preprint arXiv:2102.01567}}
  (\bibinfo{year}{2021}).
\newblock


\bibitem[\protect\citeauthoryear{Cobbe, Hilton, Klimov, and Schulman}{Cobbe
  et~al\mbox{.}}{2020}]%
        {cobbe2020phasic}
\bibfield{author}{\bibinfo{person}{Karl Cobbe}, \bibinfo{person}{Jacob Hilton},
  \bibinfo{person}{Oleg Klimov}, {and} \bibinfo{person}{John Schulman}.}
  \bibinfo{year}{2020}\natexlab{}.
\newblock \showarticletitle{Phasic policy gradient}.
\newblock \bibinfo{journal}{\emph{arXiv preprint arXiv:2009.04416}}
  (\bibinfo{year}{2020}).
\newblock


\bibitem[\protect\citeauthoryear{DASH}{DASH}{2019}]%
        {dash}
\bibfield{author}{\bibinfo{person}{DASH}.} \bibinfo{year}{2019}\natexlab{}.
\newblock \bibinfo{title}{DASH}.
\newblock
\newblock
\urldef\tempurl%
\url{https://dashif.org/}
\showURL{%
\tempurl}


\bibitem[\protect\citeauthoryear{Du, Zhuo, Li, Zhang, Li, and Zhang}{Du
  et~al\mbox{.}}{2020}]%
        {du2020video}
\bibfield{author}{\bibinfo{person}{Lina Du}, \bibinfo{person}{Li Zhuo},
  \bibinfo{person}{Jiafeng Li}, \bibinfo{person}{Jing Zhang},
  \bibinfo{person}{Xiaoguang Li}, {and} \bibinfo{person}{Hui Zhang}.}
  \bibinfo{year}{2020}\natexlab{}.
\newblock \showarticletitle{Video quality of experience metric for dynamic
  adaptive streaming services using DASH standard and deep spatial-temporal
  representation of video}.
\newblock \bibinfo{journal}{\emph{Applied Sciences}} \bibinfo{volume}{10},
  \bibinfo{number}{5} (\bibinfo{year}{2020}), \bibinfo{pages}{1793}.
\newblock


\bibitem[\protect\citeauthoryear{Duanmu, Liu, Chen, Li, Wang, Wang, and
  Gao}{Duanmu et~al\mbox{.}}{2019}]%
        {duanmu2019knowledge}
\bibfield{author}{\bibinfo{person}{Zhengfang Duanmu}, \bibinfo{person}{Wentao
  Liu}, \bibinfo{person}{Diqi Chen}, \bibinfo{person}{Zhuoran Li},
  \bibinfo{person}{Zhou Wang}, \bibinfo{person}{Yizhou Wang}, {and}
  \bibinfo{person}{Wen Gao}.} \bibinfo{year}{2019}\natexlab{}.
\newblock \showarticletitle{A knowledge-driven quality-of-experience model for
  adaptive streaming videos}.
\newblock \bibinfo{journal}{\emph{arXiv preprint arXiv:1911.07944}}
  (\bibinfo{year}{2019}).
\newblock


\bibitem[\protect\citeauthoryear{Duanmu, Liu, Li, Chen, Wang, Wang, and
  Gao}{Duanmu et~al\mbox{.}}{2020}]%
        {duanmu2020assessing}
\bibfield{author}{\bibinfo{person}{Zhengfang Duanmu}, \bibinfo{person}{Wentao
  Liu}, \bibinfo{person}{Zhuoran Li}, \bibinfo{person}{Diqi Chen},
  \bibinfo{person}{Zhou Wang}, \bibinfo{person}{Yizhou Wang}, {and}
  \bibinfo{person}{Wen Gao}.} \bibinfo{year}{2020}\natexlab{}.
\newblock \showarticletitle{Assessing the quality-of-experience of adaptive
  bitrate video streaming}.
\newblock \bibinfo{journal}{\emph{arXiv preprint arXiv:2008.08804}}
  (\bibinfo{year}{2020}).
\newblock


\bibitem[\protect\citeauthoryear{Duanmu, Rehman, and Wang}{Duanmu
  et~al\mbox{.}}{2018}]%
        {duanmu2018quality}
\bibfield{author}{\bibinfo{person}{Zhengfang Duanmu}, \bibinfo{person}{Abdul
  Rehman}, {and} \bibinfo{person}{Zhou Wang}.} \bibinfo{year}{2018}\natexlab{}.
\newblock \showarticletitle{A quality-of-experience database for adaptive video
  streaming}.
\newblock \bibinfo{journal}{\emph{IEEE Transactions on Broadcasting}}
  \bibinfo{volume}{64}, \bibinfo{number}{2} (\bibinfo{year}{2018}),
  \bibinfo{pages}{474--487}.
\newblock


\bibitem[\protect\citeauthoryear{Feng, Sun, Qi, Wang, and Liao}{Feng
  et~al\mbox{.}}{2019}]%
        {feng2019vabis}
\bibfield{author}{\bibinfo{person}{Tongtong Feng}, \bibinfo{person}{Haifeng
  Sun}, \bibinfo{person}{Qi Qi}, \bibinfo{person}{Jingyu Wang}, {and}
  \bibinfo{person}{Jianxin Liao}.} \bibinfo{year}{2019}\natexlab{}.
\newblock \showarticletitle{Vabis: Video Adaptation Bitrate System for
  Time-Critical Live Streaming}.
\newblock \bibinfo{journal}{\emph{IEEE Transactions on Multimedia}}
  \bibinfo{volume}{22}, \bibinfo{number}{11} (\bibinfo{year}{2019}),
  \bibinfo{pages}{2963--2976}.
\newblock


\bibitem[\protect\citeauthoryear{Garivier and Moulines}{Garivier and
  Moulines}{2011}]%
        {garivier2011upper}
\bibfield{author}{\bibinfo{person}{Aur{\'e}lien Garivier} {and}
  \bibinfo{person}{Eric Moulines}.} \bibinfo{year}{2011}\natexlab{}.
\newblock \showarticletitle{On upper-confidence bound policies for switching
  bandit problems}. In \bibinfo{booktitle}{\emph{Algorithmic Learning Theory:
  22nd International Conference, ALT 2011, Espoo, Finland, October 5-7, 2011.
  Proceedings 22}}. Springer, \bibinfo{pages}{174--188}.
\newblock


\bibitem[\protect\citeauthoryear{group: Puffer treats SSIM$_{db}$ as
  the~quality metric.}{group: Puffer treats SSIM$_{db}$ as the~quality
  metric.}{2023}]%
        {puffer2022ssimdb}
\bibfield{author}{\bibinfo{person}{Puffer group: Puffer treats SSIM$_{db}$ as
  the~quality metric.}} \bibinfo{year}{2023}\natexlab{}.
\newblock \bibinfo{title}{Puffer official repositories}.
\newblock
  \bibinfo{howpublished}{\url{https://github.com/StanfordSNR/puffer/blob/ef90af9a50543126b53720657ab9a4b16bdaa0ca/src/abr/mpc_search.cc\#L159}}.
\newblock


\bibitem[\protect\citeauthoryear{Huang, He, Gao, Deng, Acero, and Heck}{Huang
  et~al\mbox{.}}{2013}]%
        {huang2013learning}
\bibfield{author}{\bibinfo{person}{Po-Sen Huang}, \bibinfo{person}{Xiaodong
  He}, \bibinfo{person}{Jianfeng Gao}, \bibinfo{person}{Li Deng},
  \bibinfo{person}{Alex Acero}, {and} \bibinfo{person}{Larry Heck}.}
  \bibinfo{year}{2013}\natexlab{}.
\newblock \showarticletitle{Learning deep structured semantic models for web
  search using clickthrough data}. In \bibinfo{booktitle}{\emph{Proceedings of
  the 22nd ACM international conference on Information \& Knowledge
  Management}}. \bibinfo{pages}{2333--2338}.
\newblock


\bibitem[\protect\citeauthoryear{Huang}{Huang}{2020}]%
        {comycovideogithub}
\bibfield{author}{\bibinfo{person}{Tianchi Huang}.}
  \bibinfo{year}{2020}\natexlab{}.
\newblock \bibinfo{title}{Comyco Video Description Dataset}.
\newblock
  \bibinfo{howpublished}{\url{https://github.com/godka/Comyco-Video-Description-Dataset/}}.
\newblock


\bibitem[\protect\citeauthoryear{Huang, Zhang, and Sun}{Huang
  et~al\mbox{.}}{2021a}]%
        {huang2021zwei}
\bibfield{author}{\bibinfo{person}{Tianchi Huang}, \bibinfo{person}{Ruixiao
  Zhang}, {and} \bibinfo{person}{Lifeng Sun}.}
  \bibinfo{year}{2021}\natexlab{a}.
\newblock \showarticletitle{Zwei: A Self-play Reinforcement Learning Framework
  for Video Transmission Services}.
\newblock \bibinfo{journal}{\emph{IEEE Transactions on Multimedia}}
  (\bibinfo{year}{2021}).
\newblock


\bibitem[\protect\citeauthoryear{Huang, Zhang, and Sun}{Huang
  et~al\mbox{.}}{2020a}]%
        {huang2020self}
\bibfield{author}{\bibinfo{person}{Tianchi Huang}, \bibinfo{person}{Rui-Xiao
  Zhang}, {and} \bibinfo{person}{Lifeng Sun}.}
  \bibinfo{year}{2020}\natexlab{a}.
\newblock \showarticletitle{Self-play reinforcement learning for video
  transmission}. In \bibinfo{booktitle}{\emph{Proceedings of the 30th ACM
  Workshop on Network and Operating Systems Support for Digital Audio and
  Video}}. \bibinfo{pages}{7--13}.
\newblock


\bibitem[\protect\citeauthoryear{Huang, Zhang, and Sun}{Huang
  et~al\mbox{.}}{2021b}]%
        {huang2021deep}
\bibfield{author}{\bibinfo{person}{Tianchi Huang}, \bibinfo{person}{Rui-Xiao
  Zhang}, {and} \bibinfo{person}{Lifeng Sun}.}
  \bibinfo{year}{2021}\natexlab{b}.
\newblock \showarticletitle{Deep reinforced bitrate ladders for adaptive video
  streaming}. In \bibinfo{booktitle}{\emph{Proceedings of the 31st ACM Workshop
  on Network and Operating Systems Support for Digital Audio and Video}}.
  \bibinfo{pages}{66--73}.
\newblock


\bibitem[\protect\citeauthoryear{Huang, Zhang, Zhou, and Sun}{Huang
  et~al\mbox{.}}{2018}]%
        {huang2018qarc}
\bibfield{author}{\bibinfo{person}{Tianchi Huang}, \bibinfo{person}{Rui-Xiao
  Zhang}, \bibinfo{person}{Chao Zhou}, {and} \bibinfo{person}{Lifeng Sun}.}
  \bibinfo{year}{2018}\natexlab{}.
\newblock \showarticletitle{Qarc: Video quality aware rate control for
  real-time video streaming based on deep reinforcement learning}. In
  \bibinfo{booktitle}{\emph{Proceedings of the 26th ACM international
  conference on Multimedia}}. \bibinfo{pages}{1208--1216}.
\newblock


\bibitem[\protect\citeauthoryear{Huang, Zhou, Yao, Zhang, Wu, Yu, and
  Sun}{Huang et~al\mbox{.}}{2020b}]%
        {huang2020quality}
\bibfield{author}{\bibinfo{person}{Tianchi Huang}, \bibinfo{person}{Chao Zhou},
  \bibinfo{person}{Xin Yao}, \bibinfo{person}{Rui-Xiao Zhang},
  \bibinfo{person}{Chenglei Wu}, \bibinfo{person}{Bing Yu}, {and}
  \bibinfo{person}{Lifeng Sun}.} \bibinfo{year}{2020}\natexlab{b}.
\newblock \showarticletitle{Quality-aware neural adaptive video streaming with
  lifelong imitation learning}.
\newblock \bibinfo{journal}{\emph{IEEE Journal on Selected Areas in
  Communications}} \bibinfo{volume}{38}, \bibinfo{number}{10}
  (\bibinfo{year}{2020}), \bibinfo{pages}{2324--2342}.
\newblock


\bibitem[\protect\citeauthoryear{Huang, Zhou, Zhang, Wu, and Sun}{Huang
  et~al\mbox{.}}{2022}]%
        {huang2022learning}
\bibfield{author}{\bibinfo{person}{Tianchi Huang}, \bibinfo{person}{Chao Zhou},
  \bibinfo{person}{Rui-Xiao Zhang}, \bibinfo{person}{Chenglei Wu}, {and}
  \bibinfo{person}{Lifeng Sun}.} \bibinfo{year}{2022}\natexlab{}.
\newblock \showarticletitle{Learning tailored adaptive bitrate algorithms to
  heterogeneous network conditions: A domain-specific priors and
  meta-reinforcement learning approach}.
\newblock \bibinfo{journal}{\emph{IEEE Journal on Selected Areas in
  Communications}} \bibinfo{volume}{40}, \bibinfo{number}{8}
  (\bibinfo{year}{2022}), \bibinfo{pages}{2485--2503}.
\newblock


\bibitem[\protect\citeauthoryear{Huang, Zhou, Zhang, Wu, Yao, and Sun}{Huang
  et~al\mbox{.}}{2019b}]%
        {huang2019comyco}
\bibfield{author}{\bibinfo{person}{Tianchi Huang}, \bibinfo{person}{Chao Zhou},
  \bibinfo{person}{Rui-Xiao Zhang}, \bibinfo{person}{Chenglei Wu},
  \bibinfo{person}{Xin Yao}, {and} \bibinfo{person}{Lifeng Sun}.}
  \bibinfo{year}{2019}\natexlab{b}.
\newblock \showarticletitle{Comyco: Quality-aware adaptive video streaming via
  imitation learning}. In \bibinfo{booktitle}{\emph{Proceedings of the 27th ACM
  International Conference on Multimedia}}. \bibinfo{pages}{429--437}.
\newblock


\bibitem[\protect\citeauthoryear{Huang, Zhou, Zhang, Wu, Yao, and Sun}{Huang
  et~al\mbox{.}}{2020c}]%
        {huang2020stick}
\bibfield{author}{\bibinfo{person}{Tianchi Huang}, \bibinfo{person}{Chao Zhou},
  \bibinfo{person}{Rui-Xiao Zhang}, \bibinfo{person}{Chenglei Wu},
  \bibinfo{person}{Xin Yao}, {and} \bibinfo{person}{Lifeng Sun}.}
  \bibinfo{year}{2020}\natexlab{c}.
\newblock \showarticletitle{Stick: A Harmonious Fusion of Buffer-based and
  Learning-based Approach for Adaptive Streaming}. In
  \bibinfo{booktitle}{\emph{IEEE INFOCOM 2020-IEEE Conference on Computer
  Communications}}. IEEE, \bibinfo{pages}{1967--1976}.
\newblock


\bibitem[\protect\citeauthoryear{Huang, Ekanadham, Berglund, and Li}{Huang
  et~al\mbox{.}}{2019a}]%
        {huang2019hindsight}
\bibfield{author}{\bibinfo{person}{Te-Yuan Huang}, \bibinfo{person}{Chaitanya
  Ekanadham}, \bibinfo{person}{Andrew~J Berglund}, {and} \bibinfo{person}{Zhi
  Li}.} \bibinfo{year}{2019}\natexlab{a}.
\newblock \showarticletitle{Hindsight: Evaluate video bitrate adaptation at
  scale}. In \bibinfo{booktitle}{\emph{Proceedings of the 10th ACM Multimedia
  Systems Conference}}. \bibinfo{pages}{86--97}.
\newblock


\bibitem[\protect\citeauthoryear{Huang, Johari, McKeown, Trunnell, and
  Watson}{Huang et~al\mbox{.}}{2014}]%
        {huang2015buffer}
\bibfield{author}{\bibinfo{person}{Te-Yuan Huang}, \bibinfo{person}{Ramesh
  Johari}, \bibinfo{person}{Nick McKeown}, \bibinfo{person}{Matthew Trunnell},
  {and} \bibinfo{person}{Mark Watson}.} \bibinfo{year}{2014}\natexlab{}.
\newblock \showarticletitle{A buffer-based approach to rate adaptation:
  Evidence from a large video streaming service}.
\newblock \bibinfo{journal}{\emph{SIGCOMM 2014}} \bibinfo{volume}{44},
  \bibinfo{number}{4} (\bibinfo{year}{2014}), \bibinfo{pages}{187--198}.
\newblock


\bibitem[\protect\citeauthoryear{Huynh-Thu, Garcia, Speranza, Corriveau, and
  Raake}{Huynh-Thu et~al\mbox{.}}{2010}]%
        {huynh2010study}
\bibfield{author}{\bibinfo{person}{Quan Huynh-Thu},
  \bibinfo{person}{Marie-Neige Garcia}, \bibinfo{person}{Filippo Speranza},
  \bibinfo{person}{Philip Corriveau}, {and} \bibinfo{person}{Alexander Raake}.}
  \bibinfo{year}{2010}\natexlab{}.
\newblock \showarticletitle{Study of rating scales for subjective quality
  assessment of high-definition video}.
\newblock \bibinfo{journal}{\emph{IEEE Transactions on Broadcasting}}
  \bibinfo{volume}{57}, \bibinfo{number}{1} (\bibinfo{year}{2010}),
  \bibinfo{pages}{1--14}.
\newblock


\bibitem[\protect\citeauthoryear{Jiang, Sekar, and Zhang}{Jiang
  et~al\mbox{.}}{2014}]%
        {jiang2014improving}
\bibfield{author}{\bibinfo{person}{Junchen Jiang}, \bibinfo{person}{Vyas
  Sekar}, {and} \bibinfo{person}{Hui Zhang}.} \bibinfo{year}{2014}\natexlab{}.
\newblock \showarticletitle{Improving fairness, efficiency, and stability in
  http-based adaptive video streaming with festive}.
\newblock \bibinfo{journal}{\emph{TON}} \bibinfo{volume}{22},
  \bibinfo{number}{1} (\bibinfo{year}{2014}), \bibinfo{pages}{326--340}.
\newblock


\bibitem[\protect\citeauthoryear{Kan, Jiang, Li, Dai, Zou, and Xiong}{Kan
  et~al\mbox{.}}{2022}]%
        {kan2022improving}
\bibfield{author}{\bibinfo{person}{Nuowen Kan}, \bibinfo{person}{Yuankun
  Jiang}, \bibinfo{person}{Chenglin Li}, \bibinfo{person}{Wenrui Dai},
  \bibinfo{person}{Junni Zou}, {and} \bibinfo{person}{Hongkai Xiong}.}
  \bibinfo{year}{2022}\natexlab{}.
\newblock \showarticletitle{Improving Generalization for Neural Adaptive Video
  Streaming via Meta Reinforcement Learning}. In
  \bibinfo{booktitle}{\emph{Proceedings of the 30th ACM International
  Conference on Multimedia}}. \bibinfo{pages}{3006--3016}.
\newblock


\bibitem[\protect\citeauthoryear{Kingma and Ba}{Kingma and Ba}{2014}]%
        {kingma2014adam}
\bibfield{author}{\bibinfo{person}{Diederik~P Kingma} {and}
  \bibinfo{person}{Jimmy Ba}.} \bibinfo{year}{2014}\natexlab{}.
\newblock \showarticletitle{Adam: A method for stochastic optimization}.
\newblock \bibinfo{journal}{\emph{arXiv preprint arXiv:1412.6980}}
  (\bibinfo{year}{2014}).
\newblock


\bibitem[\protect\citeauthoryear{Li, Koyamada, Ye, Liu, Wang, Yang, Zhao, Qin,
  Liu, and Hon}{Li et~al\mbox{.}}{2020}]%
        {li2020suphx}
\bibfield{author}{\bibinfo{person}{Junjie Li}, \bibinfo{person}{Sotetsu
  Koyamada}, \bibinfo{person}{Qiwei Ye}, \bibinfo{person}{Guoqing Liu},
  \bibinfo{person}{Chao Wang}, \bibinfo{person}{Ruihan Yang},
  \bibinfo{person}{Li Zhao}, \bibinfo{person}{Tao Qin},
  \bibinfo{person}{Tie-Yan Liu}, {and} \bibinfo{person}{Hsiao-Wuen Hon}.}
  \bibinfo{year}{2020}\natexlab{}.
\newblock \showarticletitle{Suphx: Mastering mahjong with deep reinforcement
  learning}.
\newblock \bibinfo{journal}{\emph{arXiv preprint arXiv:2003.13590}}
  (\bibinfo{year}{2020}).
\newblock


\bibitem[\protect\citeauthoryear{Li, Huang, Wang, Wu, Liu, and Wang}{Li
  et~al\mbox{.}}{2022}]%
        {li2022apprenticeship}
\bibfield{author}{\bibinfo{person}{Weihe Li}, \bibinfo{person}{Jiawei Huang},
  \bibinfo{person}{Shiqi Wang}, \bibinfo{person}{Chuliang Wu},
  \bibinfo{person}{Sen Liu}, {and} \bibinfo{person}{Jianxin Wang}.}
  \bibinfo{year}{2022}\natexlab{}.
\newblock \showarticletitle{An Apprenticeship Learning Approach for Adaptive
  Video Streaming Based on Chunk Quality and User Preference}.
\newblock \bibinfo{journal}{\emph{IEEE Transactions on Multimedia}}
  (\bibinfo{year}{2022}).
\newblock


\bibitem[\protect\citeauthoryear{Li, Zhang, Chen, and Ma}{Li
  et~al\mbox{.}}{2023}]%
        {li2023mamba}
\bibfield{author}{\bibinfo{person}{Yueheng Li}, \bibinfo{person}{Zicheng
  Zhang}, \bibinfo{person}{Hao Chen}, {and} \bibinfo{person}{Zhan Ma}.}
  \bibinfo{year}{2023}\natexlab{}.
\newblock \showarticletitle{Mamba: Bringing Multi-Dimensional ABR to WebRTC}.
  In \bibinfo{booktitle}{\emph{Proceedings of the 31st ACM International
  Conference on Multimedia}}.
\newblock


\bibitem[\protect\citeauthoryear{Li, Zhu, Gahm, Pan, Hu, Begen, and Oran}{Li
  et~al\mbox{.}}{2014}]%
        {li2014probe}
\bibfield{author}{\bibinfo{person}{Zhi Li}, \bibinfo{person}{Xiaoqing Zhu},
  \bibinfo{person}{Joshua Gahm}, \bibinfo{person}{Rong Pan},
  \bibinfo{person}{Hao Hu}, \bibinfo{person}{Ali~C Begen}, {and}
  \bibinfo{person}{David Oran}.} \bibinfo{year}{2014}\natexlab{}.
\newblock \showarticletitle{Probe and adapt: Rate adaptation for HTTP video
  streaming at scale}.
\newblock \bibinfo{journal}{\emph{IEEE JASC}} \bibinfo{volume}{32},
  \bibinfo{number}{4} (\bibinfo{year}{2014}), \bibinfo{pages}{719--733}.
\newblock


\bibitem[\protect\citeauthoryear{Lin, Feng, Santos, Yu, Xiang, Zhou, and
  Bengio}{Lin et~al\mbox{.}}{2017}]%
        {lin2017structured}
\bibfield{author}{\bibinfo{person}{Zhouhan Lin}, \bibinfo{person}{Minwei Feng},
  \bibinfo{person}{Cicero Nogueira~dos Santos}, \bibinfo{person}{Mo Yu},
  \bibinfo{person}{Bing Xiang}, \bibinfo{person}{Bowen Zhou}, {and}
  \bibinfo{person}{Yoshua Bengio}.} \bibinfo{year}{2017}\natexlab{}.
\newblock \showarticletitle{A structured self-attentive sentence embedding}.
\newblock \bibinfo{journal}{\emph{arXiv preprint arXiv:1703.03130}}
  (\bibinfo{year}{2017}).
\newblock


\bibitem[\protect\citeauthoryear{Lv, Qinghua, Wang, Li, and Xie}{Lv
  et~al\mbox{.}}{2022}]%
        {lv2022lumos}
\bibfield{author}{\bibinfo{person}{Gerui Lv}, \bibinfo{person}{Wu Qinghua},
  \bibinfo{person}{Weiran Wang}, \bibinfo{person}{Zhenyu Li}, {and}
  \bibinfo{person}{Gaogang Xie}.} \bibinfo{year}{2022}\natexlab{}.
\newblock \showarticletitle{Lumos: towards Better Video Streaming QoE through
  Accurate Throughput Prediction}. In \bibinfo{booktitle}{\emph{IEEE INFOCOM
  2022-IEEE Conference on Computer Communications}}. IEEE,
  \bibinfo{pages}{1--10}.
\newblock


\bibitem[\protect\citeauthoryear{Mao, Negi, Narayan, Wang, Yang, Wang, Marcus,
  Addanki, Khani, He, et~al\mbox{.}}{Mao et~al\mbox{.}}{2019}]%
        {mao2019park}
\bibfield{author}{\bibinfo{person}{Hongzi Mao}, \bibinfo{person}{Parimarjan
  Negi}, \bibinfo{person}{Akshay Narayan}, \bibinfo{person}{Hanrui Wang},
  \bibinfo{person}{Jiacheng Yang}, \bibinfo{person}{Haonan Wang},
  \bibinfo{person}{Ryan Marcus}, \bibinfo{person}{Ravichandra Addanki},
  \bibinfo{person}{Mehrdad Khani}, \bibinfo{person}{Songtao He},
  {et~al\mbox{.}}} \bibinfo{year}{2019}\natexlab{}.
\newblock \showarticletitle{Park: An open platform for learning augmented
  computer systems}. In \bibinfo{booktitle}{\emph{NIPS 2019}}.
\newblock


\bibitem[\protect\citeauthoryear{Mao, Netravali, Alizadeh, et~al\mbox{.}}{Mao
  et~al\mbox{.}}{2017}]%
        {mao2017neural}
\bibfield{author}{\bibinfo{person}{Hongzi Mao}, \bibinfo{person}{Ravi
  Netravali}, \bibinfo{person}{Mohammad Alizadeh}, {et~al\mbox{.}}}
  \bibinfo{year}{2017}\natexlab{}.
\newblock \showarticletitle{Neural adaptive video streaming with pensieve}. In
  \bibinfo{booktitle}{\emph{SIGCOMM 2017}}. ACM, \bibinfo{pages}{197--210}.
\newblock


\bibitem[\protect\citeauthoryear{Meng, Guo, Shen, Chen, Zhou, Wang, Zhang, Xu,
  Sun, and Hu}{Meng et~al\mbox{.}}{2021}]%
        {meng2021practically}
\bibfield{author}{\bibinfo{person}{Zili Meng}, \bibinfo{person}{Yaning Guo},
  \bibinfo{person}{Yixin Shen}, \bibinfo{person}{Jing Chen},
  \bibinfo{person}{Chao Zhou}, \bibinfo{person}{Minhu Wang},
  \bibinfo{person}{Jia Zhang}, \bibinfo{person}{Mingwei Xu},
  \bibinfo{person}{Chen Sun}, {and} \bibinfo{person}{Hongxin Hu}.}
  \bibinfo{year}{2021}\natexlab{}.
\newblock \showarticletitle{Practically Deploying Heavyweight Adaptive Bitrate
  Algorithms With Teacher-Student Learning}.
\newblock \bibinfo{journal}{\emph{IEEE/ACM Transactions on Networking}}
  \bibinfo{volume}{29}, \bibinfo{number}{2} (\bibinfo{year}{2021}),
  \bibinfo{pages}{723--736}.
\newblock


\bibitem[\protect\citeauthoryear{Narayanan, Ramadan, Mehta, Hu, Liu, Fezeu,
  Dayalan, Verma, Ji, Li, Qian, and Zhang}{Narayanan et~al\mbox{.}}{2020}]%
        {Narayanan2020lumos5g}
\bibfield{author}{\bibinfo{person}{Arvind Narayanan}, \bibinfo{person}{Eman
  Ramadan}, \bibinfo{person}{Rishabh Mehta}, \bibinfo{person}{Xinyue Hu},
  \bibinfo{person}{Qingxu Liu}, \bibinfo{person}{Rostand A.~K. Fezeu},
  \bibinfo{person}{Udhaya~Kumar Dayalan}, \bibinfo{person}{Saurabh Verma},
  \bibinfo{person}{Peiqi Ji}, \bibinfo{person}{Tao Li}, \bibinfo{person}{Feng
  Qian}, {and} \bibinfo{person}{Zhi-Li Zhang}.}
  \bibinfo{year}{2020}\natexlab{}.
\newblock \showarticletitle{Lumos5G: Mapping and Predicting Commercial MmWave
  5G Throughput}. In \bibinfo{booktitle}{\emph{Proceedings of the ACM Internet
  Measurement Conference}} (Virtual Event, USA) \emph{(\bibinfo{series}{IMC
  '20})}. \bibinfo{publisher}{Association for Computing Machinery},
  \bibinfo{address}{New York, NY, USA}, \bibinfo{pages}{176–193}.
\newblock
\showISBNx{9781450381383}
\urldef\tempurl%
\url{https://doi.org/10.1145/3419394.3423629}
\showDOI{\tempurl}


\bibitem[\protect\citeauthoryear{Narayanan, Zhang, Zhu, Hassan, Jin, Zhu,
  Zhang, Rybkin, Yang, Mao, et~al\mbox{.}}{Narayanan et~al\mbox{.}}{2021}]%
        {narayanan2021variegated}
\bibfield{author}{\bibinfo{person}{Arvind Narayanan}, \bibinfo{person}{Xumiao
  Zhang}, \bibinfo{person}{Ruiyang Zhu}, \bibinfo{person}{Ahmad Hassan},
  \bibinfo{person}{Shuowei Jin}, \bibinfo{person}{Xiao Zhu},
  \bibinfo{person}{Xiaoxuan Zhang}, \bibinfo{person}{Denis Rybkin},
  \bibinfo{person}{Zhengxuan Yang}, \bibinfo{person}{Zhuoqing~Morley Mao},
  {et~al\mbox{.}}} \bibinfo{year}{2021}\natexlab{}.
\newblock \showarticletitle{A variegated look at 5G in the wild: performance,
  power, and QoE implications}. In \bibinfo{booktitle}{\emph{Proceedings of the
  2021 ACM SIGCOMM 2021 Conference}}. \bibinfo{pages}{610--625}.
\newblock


\bibitem[\protect\citeauthoryear{OpenAI}{OpenAI}{2022}]%
        {chatgpt2023}
\bibfield{author}{\bibinfo{person}{OpenAI}.} \bibinfo{year}{2022}\natexlab{}.
\newblock \bibinfo{title}{Introducing ChatGPT}.
\newblock
\newblock
\urldef\tempurl%
\url{https://openai.com/blog/chatgpt#OpenAI}
\showURL{%
\tempurl}


\bibitem[\protect\citeauthoryear{Ouyang, Wu, Jiang, Almeida, Wainwright,
  Mishkin, Zhang, Agarwal, Slama, Ray, et~al\mbox{.}}{Ouyang
  et~al\mbox{.}}{2022}]%
        {ouyang2022training}
\bibfield{author}{\bibinfo{person}{Long Ouyang}, \bibinfo{person}{Jeffrey Wu},
  \bibinfo{person}{Xu Jiang}, \bibinfo{person}{Diogo Almeida},
  \bibinfo{person}{Carroll Wainwright}, \bibinfo{person}{Pamela Mishkin},
  \bibinfo{person}{Chong Zhang}, \bibinfo{person}{Sandhini Agarwal},
  \bibinfo{person}{Katarina Slama}, \bibinfo{person}{Alex Ray},
  {et~al\mbox{.}}} \bibinfo{year}{2022}\natexlab{}.
\newblock \showarticletitle{Training language models to follow instructions
  with human feedback}.
\newblock \bibinfo{journal}{\emph{Advances in Neural Information Processing
  Systems}}  \bibinfo{volume}{35} (\bibinfo{year}{2022}),
  \bibinfo{pages}{27730--27744}.
\newblock


\bibitem[\protect\citeauthoryear{Pereira, Schmidt, and Herfet}{Pereira
  et~al\mbox{.}}{2018}]%
        {pereira2018cross}
\bibfield{author}{\bibinfo{person}{Pablo~Gil Pereira}, \bibinfo{person}{Andreas
  Schmidt}, {and} \bibinfo{person}{Thorsten Herfet}.}
  \bibinfo{year}{2018}\natexlab{}.
\newblock \showarticletitle{Cross-layer effects on training neural algorithms
  for video streaming}. In \bibinfo{booktitle}{\emph{Proceedings of the 28th
  ACM SIGMM Workshop on Network and Operating Systems Support for Digital Audio
  and Video}}. \bibinfo{pages}{43--48}.
\newblock


\bibitem[\protect\citeauthoryear{Qiao, Wang, and Liu}{Qiao
  et~al\mbox{.}}{2020}]%
        {qiao2020beyond}
\bibfield{author}{\bibinfo{person}{Chunyu Qiao}, \bibinfo{person}{Jiliang
  Wang}, {and} \bibinfo{person}{Yunhao Liu}.} \bibinfo{year}{2020}\natexlab{}.
\newblock \showarticletitle{Beyond QoE: Diversity adaptation in video streaming
  at the edge}.
\newblock \bibinfo{journal}{\emph{IEEE/ACM Transactions on Networking}}
  \bibinfo{volume}{29}, \bibinfo{number}{1} (\bibinfo{year}{2020}),
  \bibinfo{pages}{289--302}.
\newblock


\bibitem[\protect\citeauthoryear{Rassool}{Rassool}{2017}]%
        {rassool2017vmaf}
\bibfield{author}{\bibinfo{person}{Reza Rassool}.}
  \bibinfo{year}{2017}\natexlab{}.
\newblock \showarticletitle{VMAF reproducibility: Validating a perceptual
  practical video quality metric}. In \bibinfo{booktitle}{\emph{Broadband
  Multimedia Systems and Broadcasting (BMSB), 2017 IEEE International Symposium
  on}}. IEEE, \bibinfo{pages}{1--2}.
\newblock


\bibitem[\protect\citeauthoryear{Rec}{Rec}{2006}]%
        {rec2006p}
\bibfield{author}{\bibinfo{person}{ITUT Rec}.} \bibinfo{year}{2006}\natexlab{}.
\newblock \showarticletitle{P. 800.1, mean opinion score (mos) terminology}.
\newblock \bibinfo{journal}{\emph{International Telecommunication Union,
  Geneva}} (\bibinfo{year}{2006}).
\newblock


\bibitem[\protect\citeauthoryear{Recommendation}{Recommendation}{2017}]%
        {recommendation20171203}
\bibfield{author}{\bibinfo{person}{ITUTP Recommendation}.}
  \bibinfo{year}{2017}\natexlab{}.
\newblock \showarticletitle{1203.3,“Parametric bitstream-based quality
  assessment of progressive download and adaptive audiovisual streaming
  services over reliable transport-Quality integration module,”}.
\newblock \bibinfo{journal}{\emph{International Telecommunication Union}}
  (\bibinfo{year}{2017}).
\newblock


\bibitem[\protect\citeauthoryear{Report}{Report}{2016}]%
        {bworld}
\bibfield{author}{\bibinfo{person}{Measuring Fixed~Broadband Report}.}
  \bibinfo{year}{2016}\natexlab{}.
\newblock \bibinfo{title}{Raw Data Measuring Broadband America 2016}.
\newblock
  \bibinfo{howpublished}{https://www.fcc.gov/reports-research/reports/measuring-broadband-america/raw-data-measuring-broadband-america-2016}.
\newblock
\newblock
\shownote{[Online; accessed 19-July-2016].}


\bibitem[\protect\citeauthoryear{Riiser, Vigmostad, Griwodz, and
  Halvorsen}{Riiser et~al\mbox{.}}{2013}]%
        {riiser2013commute}
\bibfield{author}{\bibinfo{person}{Haakon Riiser}, \bibinfo{person}{Paul
  Vigmostad}, \bibinfo{person}{Carsten Griwodz}, {and} \bibinfo{person}{P{\aa}l
  Halvorsen}.} \bibinfo{year}{2013}\natexlab{}.
\newblock \showarticletitle{Commute path bandwidth traces from 3G networks:
  analysis and applications}. In \bibinfo{booktitle}{\emph{Proceedings of the
  4th ACM Multimedia Systems Conference}}. ACM, \bibinfo{pages}{114--118}.
\newblock


\bibitem[\protect\citeauthoryear{SANDVINE2022}{SANDVINE2022}{2023}]%
        {SANDVINE2022}
\bibfield{author}{\bibinfo{person}{SANDVINE2022}.}
  \bibinfo{year}{2023}\natexlab{}.
\newblock \bibinfo{title}{The 2022 Internet Phenomena Spotlight Report}.
\newblock \bibinfo{howpublished}{\url{https://www.sandvine.com/phenomena}}.
\newblock


\bibitem[\protect\citeauthoryear{Schulman, Moritz, Levine, Jordan, and
  Abbeel}{Schulman et~al\mbox{.}}{2015}]%
        {schulman2015high}
\bibfield{author}{\bibinfo{person}{John Schulman}, \bibinfo{person}{Philipp
  Moritz}, \bibinfo{person}{Sergey Levine}, \bibinfo{person}{Michael Jordan},
  {and} \bibinfo{person}{Pieter Abbeel}.} \bibinfo{year}{2015}\natexlab{}.
\newblock \showarticletitle{High-dimensional continuous control using
  generalized advantage estimation}.
\newblock \bibinfo{journal}{\emph{arXiv preprint arXiv:1506.02438}}
  (\bibinfo{year}{2015}).
\newblock


\bibitem[\protect\citeauthoryear{Schulman, Wolski, Dhariwal, Radford, and
  Klimov}{Schulman et~al\mbox{.}}{2017}]%
        {schulman2017proximal}
\bibfield{author}{\bibinfo{person}{John Schulman}, \bibinfo{person}{Filip
  Wolski}, \bibinfo{person}{Prafulla Dhariwal}, \bibinfo{person}{Alec Radford},
  {and} \bibinfo{person}{Oleg Klimov}.} \bibinfo{year}{2017}\natexlab{}.
\newblock \showarticletitle{Proximal policy optimization algorithms}.
\newblock \bibinfo{journal}{\emph{arXiv preprint arXiv:1707.06347}}
  (\bibinfo{year}{2017}).
\newblock


\bibitem[\protect\citeauthoryear{Spiteri, Urgaonkar, and Sitaraman}{Spiteri
  et~al\mbox{.}}{2016}]%
        {spiteri2016bola}
\bibfield{author}{\bibinfo{person}{Kevin Spiteri}, \bibinfo{person}{Rahul
  Urgaonkar}, {and} \bibinfo{person}{Ramesh~K Sitaraman}.}
  \bibinfo{year}{2016}\natexlab{}.
\newblock \showarticletitle{BOLA: Near-optimal bitrate adaptation for online
  videos}. In \bibinfo{booktitle}{\emph{INFOCOM 2016, IEEE}}. IEEE,
  \bibinfo{pages}{1--9}.
\newblock


\bibitem[\protect\citeauthoryear{Spiteri, Urgaonkar, and Sitaraman}{Spiteri
  et~al\mbox{.}}{2020}]%
        {spiteri2020bola}
\bibfield{author}{\bibinfo{person}{Kevin Spiteri}, \bibinfo{person}{Rahul
  Urgaonkar}, {and} \bibinfo{person}{Ramesh~K Sitaraman}.}
  \bibinfo{year}{2020}\natexlab{}.
\newblock \showarticletitle{BOLA: Near-optimal bitrate adaptation for online
  videos}.
\newblock \bibinfo{journal}{\emph{IEEE/ACM Transactions on Networking}}
  \bibinfo{volume}{28}, \bibinfo{number}{4} (\bibinfo{year}{2020}),
  \bibinfo{pages}{1698--1711}.
\newblock


\bibitem[\protect\citeauthoryear{Sun, Yin, Jiang, et~al\mbox{.}}{Sun
  et~al\mbox{.}}{2016}]%
        {sun2016cs2p}
\bibfield{author}{\bibinfo{person}{Yi Sun}, \bibinfo{person}{Xiaoqi Yin},
  \bibinfo{person}{Junchen Jiang}, {et~al\mbox{.}}}
  \bibinfo{year}{2016}\natexlab{}.
\newblock \showarticletitle{CS2P: Improving video bitrate selection and
  adaptation with data-driven throughput prediction}. In
  \bibinfo{booktitle}{\emph{SIGCOMM 2016}}. ACM, \bibinfo{pages}{272--285}.
\newblock


\bibitem[\protect\citeauthoryear{Sutton and Barto}{Sutton and Barto}{1998}]%
        {sutton1998reinforcement}
\bibfield{author}{\bibinfo{person}{Richard~S Sutton} {and}
  \bibinfo{person}{Andrew~G Barto}.} \bibinfo{year}{1998}\natexlab{}.
\newblock \bibinfo{booktitle}{\emph{Reinforcement learning: An introduction}}.
  Vol.~\bibinfo{volume}{1}.
\newblock \bibinfo{publisher}{MIT press Cambridge}.
\newblock


\bibitem[\protect\citeauthoryear{Sutton and Barto}{Sutton and Barto}{2018}]%
        {sutton2018reinforcement}
\bibfield{author}{\bibinfo{person}{Richard~S Sutton} {and}
  \bibinfo{person}{Andrew~G Barto}.} \bibinfo{year}{2018}\natexlab{}.
\newblock \bibinfo{booktitle}{\emph{Reinforcement learning: An introduction}}.
\newblock \bibinfo{publisher}{MIT press}.
\newblock


\bibitem[\protect\citeauthoryear{Turkkan, Dai, Raman, Kosar, Chen, Fatih~Bulut,
  Zola, and Sow}{Turkkan et~al\mbox{.}}{2022}]%
        {turkkan2022greenabr}
\bibfield{author}{\bibinfo{person}{Bekir Turkkan}, \bibinfo{person}{Ting Dai},
  \bibinfo{person}{Adithya Raman}, \bibinfo{person}{Tevfik Kosar},
  \bibinfo{person}{Changyou Chen}, \bibinfo{person}{Muhammed Fatih~Bulut},
  \bibinfo{person}{Jaroslaw Zola}, {and} \bibinfo{person}{Daby Sow}.}
  \bibinfo{year}{2022}\natexlab{}.
\newblock \showarticletitle{GreenABR: Energy-Aware Adaptive Bitrate Streaming
  with Deep Reinforcement Learning}. In \bibinfo{booktitle}{\emph{Proceedings
  of the 13th ACM Multimedia Systems Conference}}.
\newblock


\bibitem[\protect\citeauthoryear{Wei, Zhou, Kwong, Yuan, Wang, Zhu, and
  Cao}{Wei et~al\mbox{.}}{2021}]%
        {wei2021reinforcement}
\bibfield{author}{\bibinfo{person}{Xuekai Wei}, \bibinfo{person}{Mingliang
  Zhou}, \bibinfo{person}{Sam Kwong}, \bibinfo{person}{Hui Yuan},
  \bibinfo{person}{Shiqi Wang}, \bibinfo{person}{Guopu Zhu}, {and}
  \bibinfo{person}{Jingchao Cao}.} \bibinfo{year}{2021}\natexlab{}.
\newblock \showarticletitle{Reinforcement learning-based QoE-oriented dynamic
  adaptive streaming framework}.
\newblock \bibinfo{journal}{\emph{Information Sciences}}  \bibinfo{volume}{569}
  (\bibinfo{year}{2021}), \bibinfo{pages}{786--803}.
\newblock


\bibitem[\protect\citeauthoryear{Wiegand, Sullivan, Bjontegaard, and
  Luthra}{Wiegand et~al\mbox{.}}{2003}]%
        {wiegand2003overview}
\bibfield{author}{\bibinfo{person}{Thomas Wiegand}, \bibinfo{person}{Gary~J
  Sullivan}, \bibinfo{person}{Gisle Bjontegaard}, {and} \bibinfo{person}{Ajay
  Luthra}.} \bibinfo{year}{2003}\natexlab{}.
\newblock \showarticletitle{Overview of the H. 264/AVC video coding standard}.
\newblock \bibinfo{journal}{\emph{IEEE Transactions on circuits and systems for
  video technology}} \bibinfo{volume}{13}, \bibinfo{number}{7}
  (\bibinfo{year}{2003}), \bibinfo{pages}{560--576}.
\newblock


\bibitem[\protect\citeauthoryear{Yadav, Shafiei, and Ooi}{Yadav
  et~al\mbox{.}}{2017}]%
        {yadav2017quetra}
\bibfield{author}{\bibinfo{person}{Praveen~Kumar Yadav}, \bibinfo{person}{Arash
  Shafiei}, {and} \bibinfo{person}{Wei~Tsang Ooi}.}
  \bibinfo{year}{2017}\natexlab{}.
\newblock \showarticletitle{Quetra: A queuing theory approach to dash rate
  adaptation}. In \bibinfo{booktitle}{\emph{Proceedings of the 25th ACM
  international conference on Multimedia}}. \bibinfo{pages}{1130--1138}.
\newblock


\bibitem[\protect\citeauthoryear{Yan, Ayers, Zhu, Fouladi, Hong, Zhang, Levis,
  and Winstein}{Yan et~al\mbox{.}}{2020}]%
        {yan2020learning}
\bibfield{author}{\bibinfo{person}{Francis~Y Yan}, \bibinfo{person}{Hudson
  Ayers}, \bibinfo{person}{Chenzhi Zhu}, \bibinfo{person}{Sadjad Fouladi},
  \bibinfo{person}{James Hong}, \bibinfo{person}{Keyi Zhang},
  \bibinfo{person}{Philip Levis}, {and} \bibinfo{person}{Keith Winstein}.}
  \bibinfo{year}{2020}\natexlab{}.
\newblock \showarticletitle{Learning in situ: a randomized experiment in video
  streaming}. In \bibinfo{booktitle}{\emph{17th $\{$USENIX$\}$ Symposium on
  Networked Systems Design and Implementation ($\{$NSDI$\}$ 20)}}.
  \bibinfo{pages}{495--511}.
\newblock


\bibitem[\protect\citeauthoryear{Ye, Liu, Sun, Shi, Zhao, Wu, Yu, Yang, Wu,
  Guo, et~al\mbox{.}}{Ye et~al\mbox{.}}{2019}]%
        {ye2019mastering}
\bibfield{author}{\bibinfo{person}{Deheng Ye}, \bibinfo{person}{Zhao Liu},
  \bibinfo{person}{Mingfei Sun}, \bibinfo{person}{Bei Shi},
  \bibinfo{person}{Peilin Zhao}, \bibinfo{person}{Hao Wu},
  \bibinfo{person}{Hongsheng Yu}, \bibinfo{person}{Shaojie Yang},
  \bibinfo{person}{Xipeng Wu}, \bibinfo{person}{Qingwei Guo}, {et~al\mbox{.}}}
  \bibinfo{year}{2019}\natexlab{}.
\newblock \showarticletitle{Mastering Complex Control in MOBA Games with Deep
  Reinforcement Learning}.
\newblock \bibinfo{journal}{\emph{arXiv preprint arXiv:1912.09729}}
  (\bibinfo{year}{2019}).
\newblock


\bibitem[\protect\citeauthoryear{Yin, Jindal, Sekar, and Sinopoli}{Yin
  et~al\mbox{.}}{2015}]%
        {yin2015control}
\bibfield{author}{\bibinfo{person}{Xiaoqi Yin}, \bibinfo{person}{Abhishek
  Jindal}, \bibinfo{person}{Vyas Sekar}, {and} \bibinfo{person}{Bruno
  Sinopoli}.} \bibinfo{year}{2015}\natexlab{}.
\newblock \showarticletitle{A control-theoretic approach for dynamic adaptive
  video streaming over HTTP}. In \bibinfo{booktitle}{\emph{SIGCOMM 2015}}. ACM,
  \bibinfo{pages}{325--338}.
\newblock


\bibitem[\protect\citeauthoryear{Zhang, Yang, Wang, Huang, Wu, Liu, and
  Sun}{Zhang et~al\mbox{.}}{2023}]%
        {zhang2023practical}
\bibfield{author}{\bibinfo{person}{Ruixiao Zhang}, \bibinfo{person}{Changpeng
  Yang}, \bibinfo{person}{Xiaochan Wang}, \bibinfo{person}{Tianchi Huang},
  \bibinfo{person}{Chenglei Wu}, \bibinfo{person}{Jiangchuan Liu}, {and}
  \bibinfo{person}{Lifeng Sun}.} \bibinfo{year}{2023}\natexlab{}.
\newblock \showarticletitle{Practical Cloud-Edge Scheduling for Large-Scale
  Crowdsourced Live Streaming}.
\newblock \bibinfo{journal}{\emph{IEEE Transactions on Parallel and Distributed
  Systems}} (\bibinfo{year}{2023}).
\newblock


\bibitem[\protect\citeauthoryear{Zhang, Ma, Huang, Li, Liu, and Sun}{Zhang
  et~al\mbox{.}}{2020}]%
        {zhang2020leveraging}
\bibfield{author}{\bibinfo{person}{Rui-Xiao Zhang}, \bibinfo{person}{Ming Ma},
  \bibinfo{person}{Tianchi Huang}, \bibinfo{person}{Hanyu Li},
  \bibinfo{person}{Jiangchuan Liu}, {and} \bibinfo{person}{Lifeng Sun}.}
  \bibinfo{year}{2020}\natexlab{}.
\newblock \showarticletitle{Leveraging QoE heterogenity for large-scale
  livecaset scheduling}. In \bibinfo{booktitle}{\emph{Proceedings of the 28th
  ACM International Conference on Multimedia}}. \bibinfo{pages}{3678--3686}.
\newblock


\bibitem[\protect\citeauthoryear{Zhang, Ou, Sen, and Jiang}{Zhang
  et~al\mbox{.}}{2021}]%
        {zhang2021sensei}
\bibfield{author}{\bibinfo{person}{Xu Zhang}, \bibinfo{person}{Yiyang Ou},
  \bibinfo{person}{Siddhartha Sen}, {and} \bibinfo{person}{Junchen Jiang}.}
  \bibinfo{year}{2021}\natexlab{}.
\newblock \showarticletitle{$\{$SENSEI$\}$: Aligning Video Streaming Quality
  with Dynamic User Sensitivity}. In \bibinfo{booktitle}{\emph{18th USENIX
  Symposium on Networked Systems Design and Implementation (NSDI 21)}}.
  \bibinfo{pages}{303--320}.
\newblock


\bibitem[\protect\citeauthoryear{Zhang, Schmitt, Chetty, Feamster, and
  Jiang}{Zhang et~al\mbox{.}}{2022}]%
        {zhang2022enabling}
\bibfield{author}{\bibinfo{person}{Xu Zhang}, \bibinfo{person}{Paul Schmitt},
  \bibinfo{person}{Marshini Chetty}, \bibinfo{person}{Nick Feamster}, {and}
  \bibinfo{person}{Junchen Jiang}.} \bibinfo{year}{2022}\natexlab{}.
\newblock \showarticletitle{Enabling Personalized Video Quality Optimization
  with VidHoc}.
\newblock \bibinfo{journal}{\emph{arXiv preprint arXiv:2211.15959}}
  (\bibinfo{year}{2022}).
\newblock


\bibitem[\protect\citeauthoryear{Zuo, Yang, Wang, and Cui}{Zuo
  et~al\mbox{.}}{2022}]%
        {zuo2022adaptive}
\bibfield{author}{\bibinfo{person}{Xutong Zuo}, \bibinfo{person}{Jiayu Yang},
  \bibinfo{person}{Mowei Wang}, {and} \bibinfo{person}{Yong Cui}.}
  \bibinfo{year}{2022}\natexlab{}.
\newblock \showarticletitle{Adaptive bitrate with user-level QOE preference for
  video streaming}. In \bibinfo{booktitle}{\emph{IEEE INFOCOM 2022-IEEE
  Conference on Computer Communications}}. IEEE, \bibinfo{pages}{1279--1288}.
\newblock


\end{thebibliography}
\newpage
\appendix
\noindent \textbf{\LARGE Appendices}
\section{Entropy-aware Deep Reinforcement Learning}
\label{sec:rlhf2}
\begin{figure}[h]
    \centering
    \includegraphics[width=0.90\linewidth]{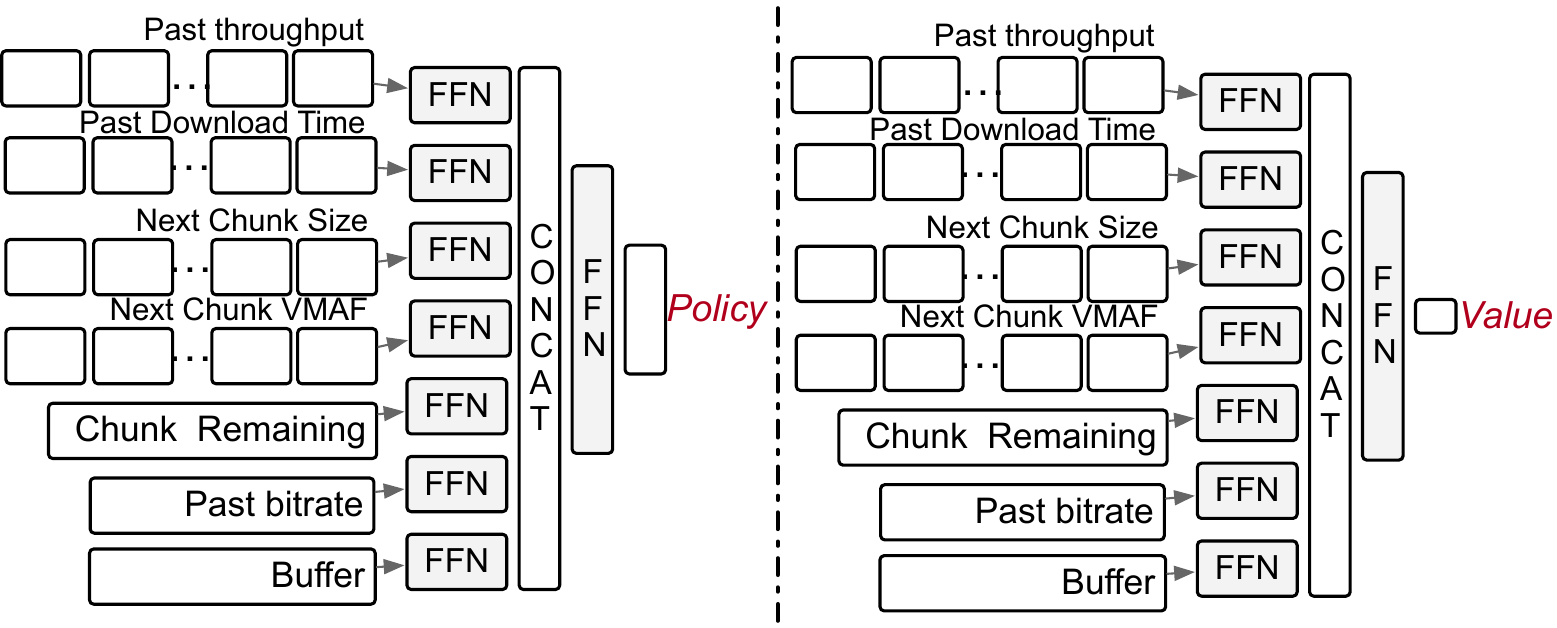}
    \caption{The NN architecture of Jade's actor network and critic network.}
    \label{fig:nn}
\end{figure}

We discuss the Dual-Clip Proximal Policy Optimization (Dual-Clip PPO) algorithm and its implementation in the \texttt{Jade} system. The detailed NN architectures are demonstrated in Figure~\ref{fig:nn}.

\subsection{Dual-Clip PPO}
We consider leveraging a novel DRL algorithm, namely Dual-Clip Proximal Policy Optimization~(Dual-Clip PPO)~\cite{ye2019mastering}. In detail, the loss function of the \texttt{Jade}'s actor-network is computed as $\mathcal{L}_\texttt{Dual}$~(Eq.~\ref{eq:policy}):

\begin{equation}
\mathcal{L}_\texttt{Dual}= \hat{\mathbb{E}_{t}}[ \mathbb{I}(\hat{A}_t < 0) \max(\mathcal{L}_\texttt{PPO}, c) \hat{A}_t + \mathbb{I}(\hat{A}_t \geq 0) \mathcal{L}_\texttt{PPO} ]
\label{eq:policy}
\end{equation}

Here $\mathbb{I}(\cdot)$ is a binary indicator function, the loss function $\mathcal{L}_\texttt{PPO}$ can be regarded as the surrogate loss function of the vanilla PPO~(Eq.\ref{eq:ppo}), which takes the probability ratio of the current policy $\pi{\phi}$ and the old policy $\pi_{{\phi}_{old}}$ into account. The advantage function $\hat{A}_t$~(Eq.\ref{eq:adv}) is learned by bootstrapping from the current estimate of the value function, $QoE^{*}$ is the smoothed QoE metric after smooth training, and $\gamma'$=0.99 is the discounted factor. It should be noted that $\hat{A}_t$ can be estimated using various other state-of-the-art methods, such as N-step TD, TD($\lambda$), V-Trace~\cite{chen2021lyapunov}, and GAE~\cite{schulman2015high}.

In summary, the Dual-clip PPO algorithm uses a double-clip approach to constrain the step size of the policy iteration and update the neural network by minimizing the \emph{clipped surrogate objective}. If the advantage function yields a value lower than zero, it will clip gradients with a lower bound of the value $\hat{A}_t$. The hyper-parameters $\epsilon$ and $c$ control how the gradient is clipped, and we set them to default values of $\epsilon=0.2$ and $c=3$~\cite{ye2019mastering}.

\begin{equation}
    \mathcal{L}_\texttt{PPO} = \min\left[\frac{\pi_{\phi}(s_t, a_t)}{\pi_{\phi_{\text{old}}}(s_t, a_t)}, \text{clip}\big(\frac{\pi_{\phi}(s_t, a_t)}{\pi_{\phi_{\text{old}}}(s_t, a_t)}, 1 \pm \epsilon\big) \right] \hat{A}_t
    \label{eq:ppo}
\end{equation}

\begin{align}
    \hat{A}_t = QoE^*_{t} + \gamma' V_{\phi_v}(s_{t+1}) - V_{\phi_v}(s_t)
    \label{eq:adv}
\end{align}

Moreover, the parameters of the \texttt{Jade}'s critic network $\phi_{v}$ are updated by minimizing the error of $\hat{A}_t$.

\begin{equation}
    \nabla \mathcal{L}_\texttt{DRL} = -\nabla_\phi \left[\mathcal{L}_\texttt{Dual} + \lambda H_\phi(s_t)\right] + \nabla_{\phi_{v}}\left[\hat{A}_t \right]^{2}.
    \label{eq:adzwei}
\end{equation}

We present the loss function $\mathcal{L}\texttt{DRL}$ in Eq.~\ref{eq:adzwei} for brevity. Furthermore, we augment the loss function with the entropy of all policies $H\phi(s_t)$, where $\lambda$ is the weight assigned to the entropy. We tune the entropy weight $\lambda$ to minimize the difference between the current entropy and the target entropy $H_{target}$. In our implementation, we set $H_{target}=0.1$~\cite{li2020suphx}.

\begin{equation}
    \lambda = \lambda + \alpha \mathbb{E}_{\{s, a\} \sim c_t}[H_{target}-H_{\pi_\phi}(s)].
\end{equation}

\subsection{Overall Learning Process}
The overall learning process is listed in Alg.~\ref{alg:rl}. It's notable that \texttt{Jade} is trained by 16 agents in parallel.

\begin{small}
\begin{algorithm}
\caption{DRL-based Training Process}
\begin{algorithmic}[1]
\Require QoE$_\texttt{lin}$: linear-based QoE model.
\Require QoE$_\texttt{DNN}$: DNN-based QoE model.
\State Initialize ABR's actor-network $\phi$, critic-network $\phi_v$.
\State Initialize smooth weight $\omega=1$.
\While {not converge}
\State \setblue{\emph{// Online trace selector}}
\State \begin{varwidth}[t]{\linewidth}
      Choose network trace $c_t$ via online trace selector: \\
      \hskip\algorithmicindent \setred{$c_t = \mathop{\argmax}_{i} \frac{\sum_{p=0}^{t} \gamma^{t-p} H_t(i) \mathbb{I}_{\{c_p=i\}} }{\sum_{p=0}^{t} \gamma^{t-p} \mathbb{I}_{\{c_p=i\}}} +  \sqrt{\frac{B\log{t}}{\sum_{p=0}^{t}\mathbb{I}_{\{c_p=i\}}}}$}.
      \end{varwidth}
\State Randomly pick a video $vid$ from the video set.
\State \setblue{\emph{// Train the policy}}
\State Initialize environment: $\texttt{Env} = \texttt{Simulator}(c_t, vid)$.
\State $D=\{\}, t=0$.
\While {not done}
\State Sample bitrate action: $a_{t} \sim \pi_\phi(s_t)$.
\State Observe next state: $s_{t+1}$ = \texttt{Env}($s_t$, $a_t$).
\State \begin{varwidth}[t]{\linewidth}
      Obtain QoE with QoE$_\texttt{lin}$ and QoE$_\texttt{DNN}$ according to: \\
      \hskip \algorithmicindent \setred{QoE$^{*}_{t} = \omega \text{QoE}_\texttt{lin}(s_{t+1}) + (1 - \omega) \text{QoE}_\texttt{DNN}(s_{t+1})$}.
      \end{varwidth}
\State $D = D \cup \{s_t, a_t, \text{QoE}^{*}_{t}\}$.
\State $t \gets t+1$
\EndWhile
\State \begin{varwidth}[t]{\linewidth}
    Train the policy $\phi$ and value $\phi_v$ using $D$ via $\mathcal{L}_\texttt{DRL}$~(Eq.\ref{eq:adzwei}): \\
    \setred{$\nabla \mathcal{L}_\texttt{DRL} = -\nabla_\phi \left[\mathcal{L}_\texttt{Dual} + \lambda H_\phi(s_t)\right] + \nabla_{\phi_{v}}\left[ \hat{A}_t \right]^{2}$}.
      \end{varwidth}
\State Update entropy weight: \setred{$\lambda = \lambda + \alpha(H_{target}-H_{\pi_\phi}(s))$}.
\State \setblue{\emph{// Enable smooth training}}
\State \begin{varwidth}[t]{\linewidth}
      Update weight $\omega$ for smooth training: \\
      \hskip\algorithmicindent \setred{$\omega = \mathbb{E}_{\{s, a\} \sim c_t}\left[\frac{-\mathop{\sum}_{a}\log \pi_\phi(s, a) \pi_\phi(s, a)}{\log |A|}\right]$}.
      \end{varwidth}
\EndWhile
\end{algorithmic}
\label{alg:rl}
\end{algorithm}
\end{small}

\begin{figure*}
    \centering
    \subfigure[VMAF vs. Stall Ratio]{
        \includegraphics[width=0.23\linewidth]{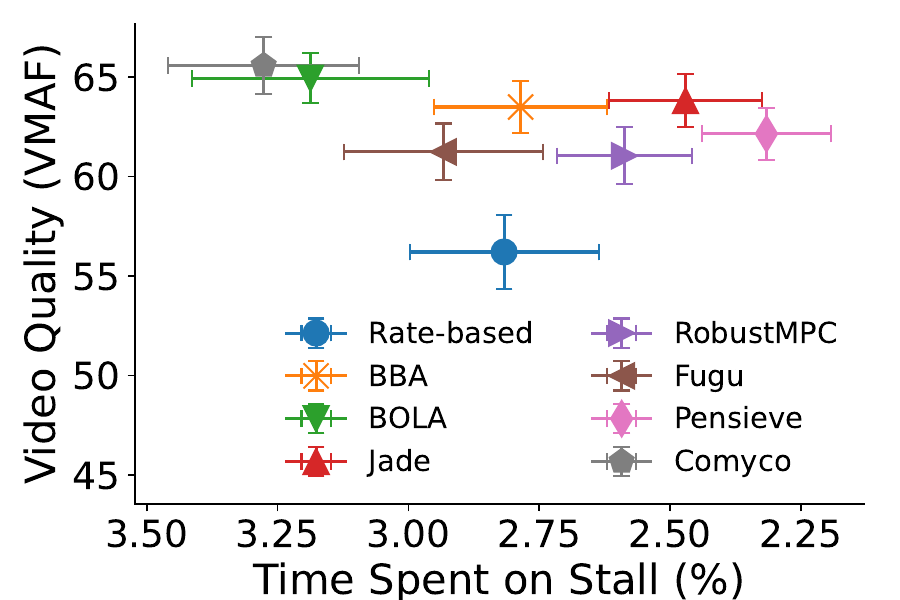}
        \label{fig:fccbr2}
    }
    \subfigure[VMAF vs. VMAF Change]{
        \includegraphics[width=0.23\linewidth]{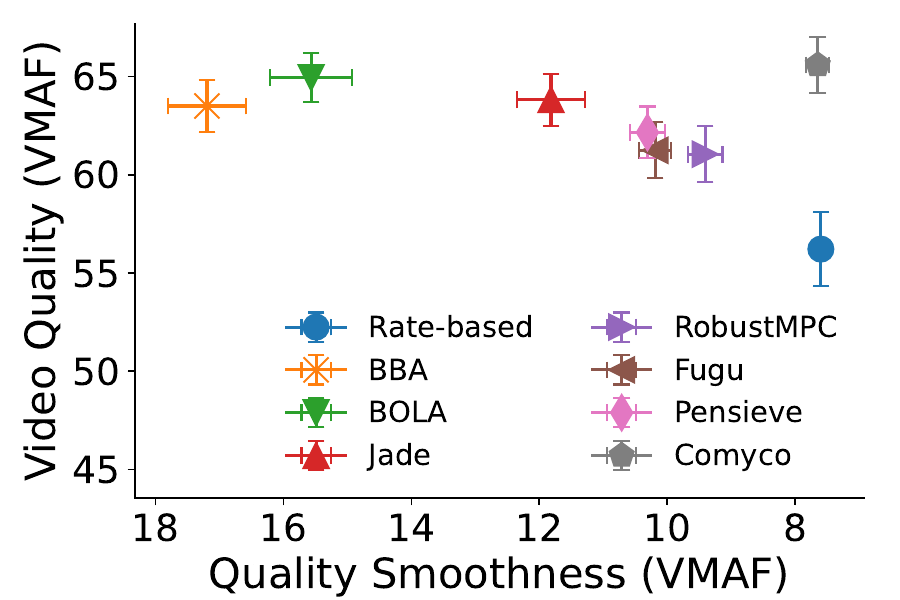}
        \label{fig:fccbs2}
    }
    \subfigure[QoE$_\texttt{DNN}$ vs. Buffer]{
        \includegraphics[width=0.23\linewidth]{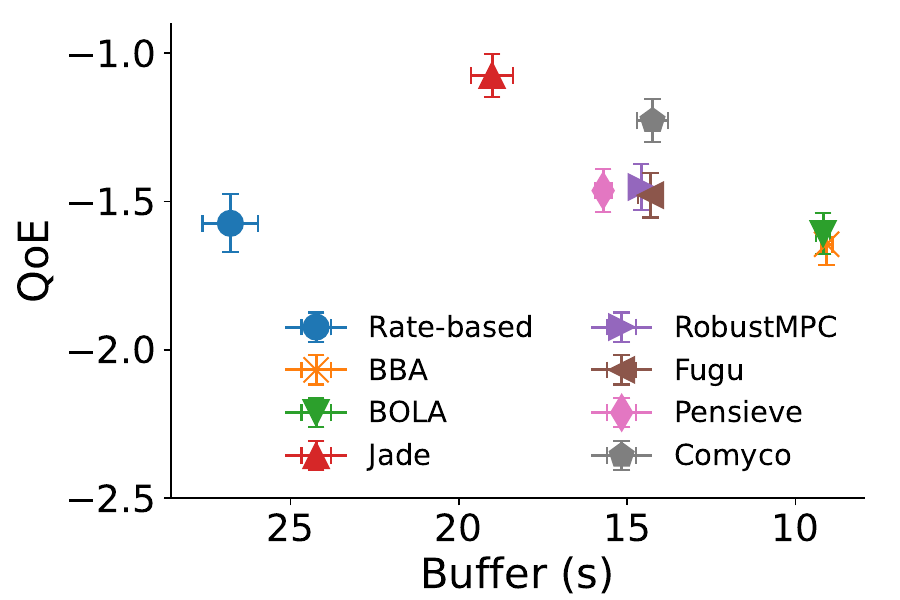}
        \label{fig:fccsr2}
    }
    \subfigure[CDF of QoE$_\texttt{DNN}$]{
        \includegraphics[width=0.23\linewidth]{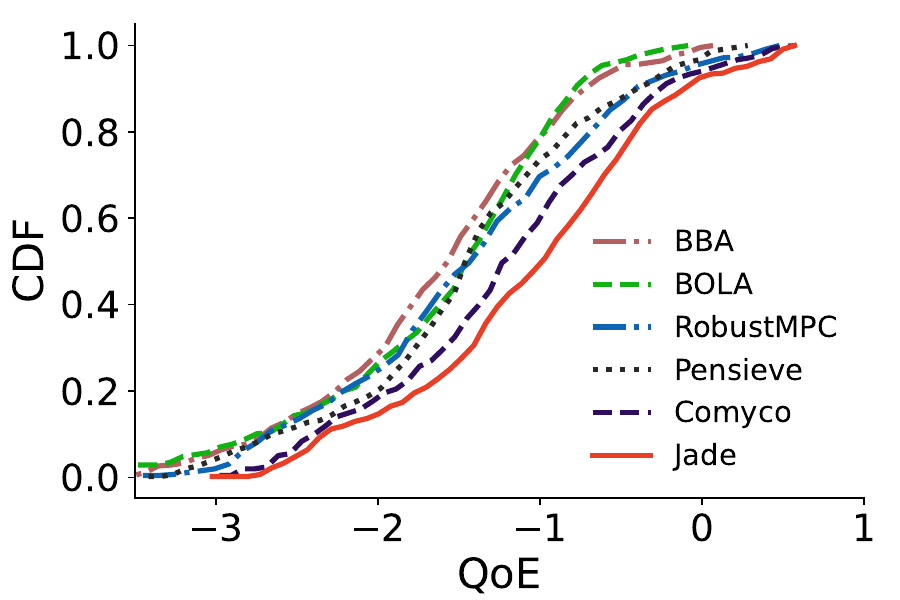}
        \label{fig:fccqoe2}
    }
    \vspace{-15pt}
    \caption{Comparing \texttt{Jade} with existing ABR algorithms using QoE$_\texttt{DNN}$. Results are collected over the FCC dataset.}
    \label{fig:fcc2}
\end{figure*}

\begin{figure*}
    \centering
    \subfigure[VMAF vs. Stall Ratio]{
        \includegraphics[width=0.23\linewidth]{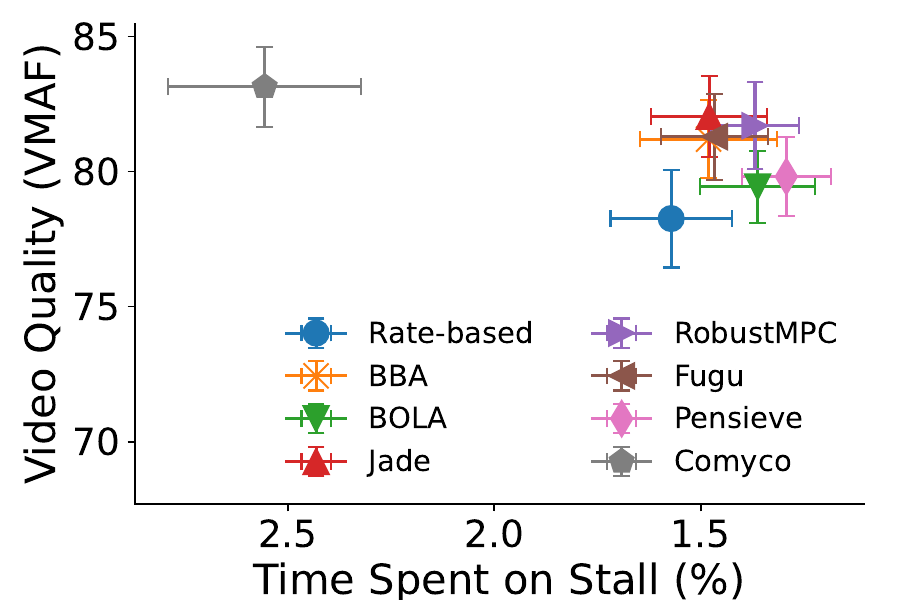}
        \label{fig:oboebr}
    }
    \subfigure[VMAF vs. VMAF Change]{
        \includegraphics[width=0.23\linewidth]{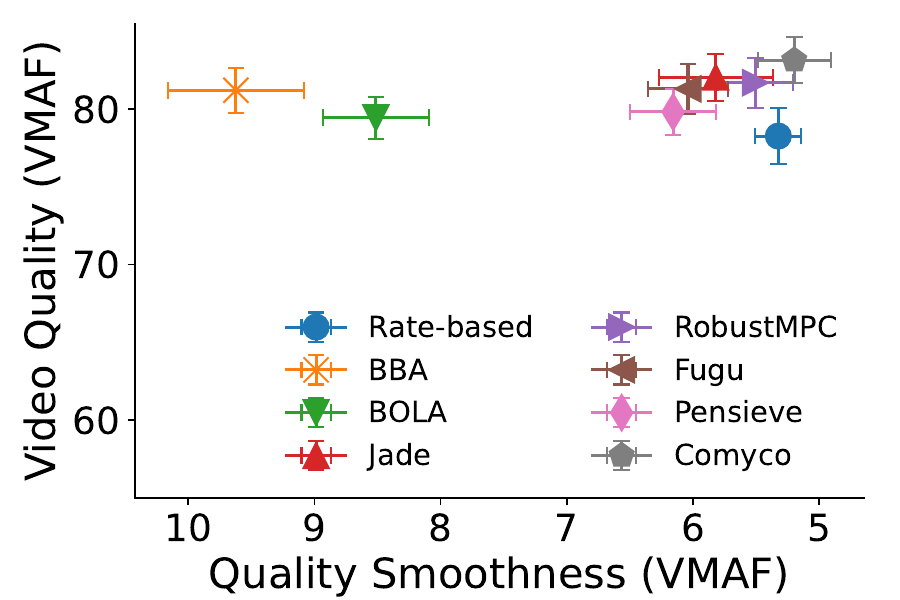}
        \label{fig:oboebs}
    }
    \subfigure[QoE$_\texttt{DNN}$ vs. Buffer]{
        \includegraphics[width=0.23\linewidth]{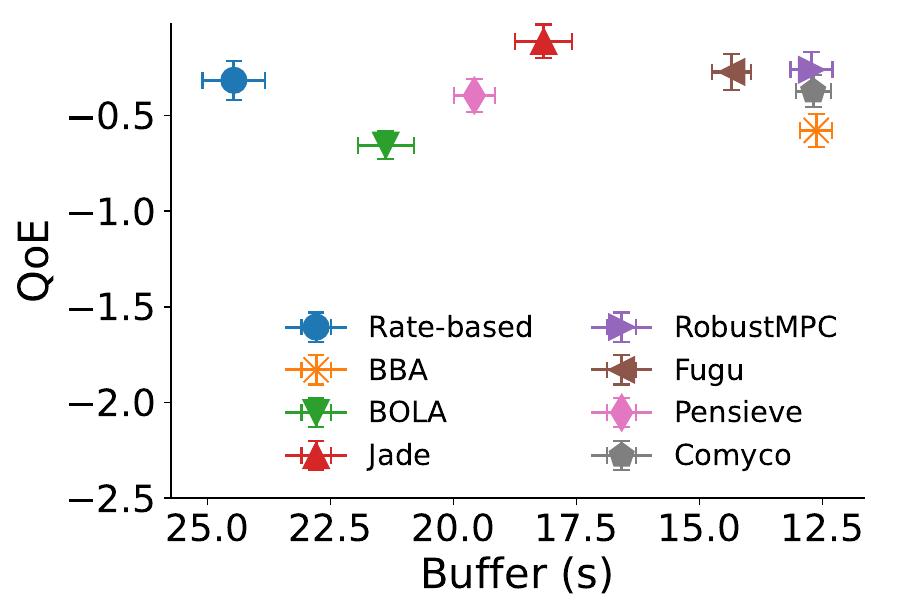}
        \label{fig:oboesr}
    }
    \subfigure[CDF of QoE$_\texttt{DNN}$]{
        \includegraphics[width=0.23\linewidth]{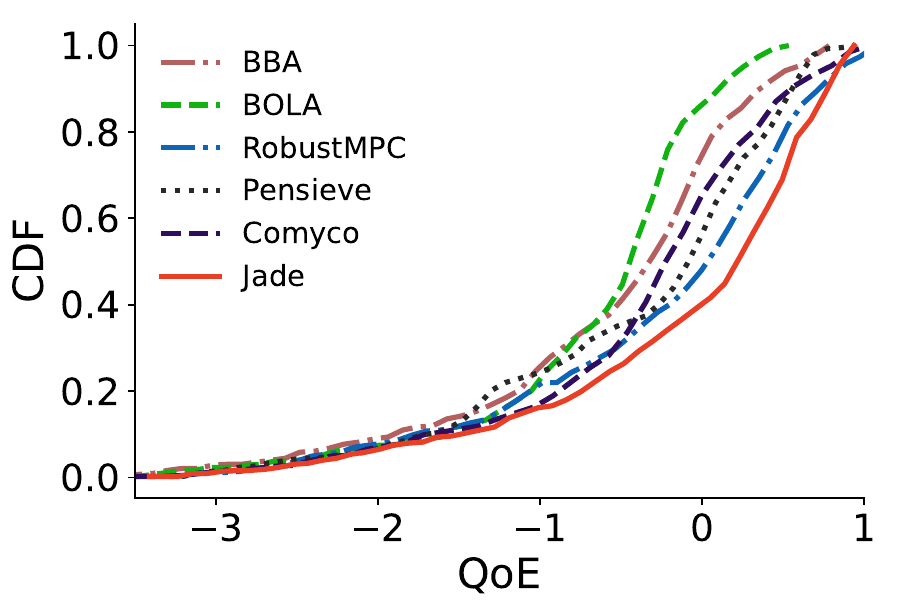}
        \label{fig:oboeqoe}
    }
    \vspace{-15pt}
    \caption{Comparing \texttt{Jade} with existing ABR algorithms using QoE$_\texttt{DNN}$. Results are collected over the Oboe dataset.}
    \label{fig:oboe}
\end{figure*}

\section{Discussion}
\subsection{Relation with Previous Paradigms}
\noindent \textbf{Relation with recent ABR algorithms.}~Existing ABR algorithms, typically learning-based schemes, often adopt deep reinforcement learning~(DRL) to learn the ABR policies without any presumptions. 
However, they are optimized with the linear-based QoE models~\cite{mao2017neural,huang2019comyco,huang2020stick,feng2019vabis,kan2022improving}. 
By contrast, \texttt{Jade}'s ABR strategy is learned by the DNN-based model, where the model often lacks generalization ability over all sessions. Thus, we propose entropy-aware DRL approaches, including smooth training methodologies and online trace selection schemes, to help tame the complexity of combining linear-based and DNN-based models.

\begin{figure}
    \centering
    \includegraphics[width=0.7\linewidth]{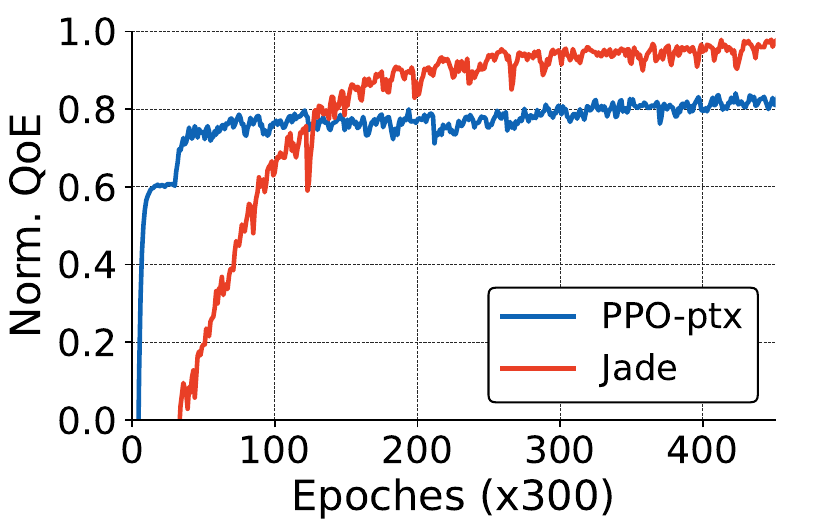}
    \vspace{-10pt}
    \caption{The learning curve of PPO-ptx.}
    \vspace{-15pt}
    \label{fig:ptx}
\end{figure}

\noindent \textbf{Relation with RLHF NLP methods.}
GPT-3 and GPT-4~\cite{ouyang2022training} have demonstrated remarkable language emergent abilities by utilizing the RLHF-based method, which is called ``PPO-ptx''. In detail, they fine-tune the Supervised fine-tuning~(SFT) model using PPO~\cite{schulman2017proximal}. Moreover, they integrate a KL penalty from the SFT model to mitigate overoptimization of the reward model and fix the performance regressions on existing NLP datasets. 
Instead, \texttt{Jade} leverages smooth training for smoothly ``sweeping'' the QoE score from QoE$_\texttt{lin}$ to QoE$_\texttt{DNN}$, with fully outlining the strengths of learning from scratch.
Figure~\ref{fig:ptx} shows the comparison results of the learning curve of PPO-ptx and \texttt{Jade}. Results are collected over the test set every 300 epochs during training. Here we observe that \texttt{Jade} outperforms PPO-ptx, with the improvements on normalized QoE of 21\%.

\section{Additional Experiments}
We report the additional experiments over FCC and Oboe datasets in Figure~\ref{fig:fcc2} and Figure~\ref{fig:oboe}.
Results illustrate the same conclusion as depicted in the main results of the paper -- \texttt{Jade} outperforms existing schemes over FCC and Oboe network traces.

\section{Parameters}
We summarize the specific parameter settings that are used throughout our experiments in Table~\ref{tbl:2} and Table~\ref{tbl:3}. We systematically categorized these datasets into low, medium, and high network paths to discuss each algorithm's practical performance and analyze the reasons behind it. Certainly, adjusting these parameters could yield improved results, while please note, the optimization of hyper-parameters is not the primary focus of our work.

\begin{table}
\caption{Parameters for \texttt{Jade}}
\begin{small}
\begin{tabular}{ccc}
\hline
Parameters & Value    & Description                   \\ \hline
$K$ & \texttt{8192} & Batch size for rank-based QoE model \\
$|A|$ & \texttt{6}~\cite{mao2017neural,huang2019comyco} & The number of bitrate levels \\
$k$ & \texttt{8}~\cite{mao2017neural,huang2019comyco} & Past $k$ metric time as a sequence \\
B & \texttt{0.2} & Hyper-parameter for online trace selector \\ \hline
$\gamma$ & \texttt{0.999} & Discounted factor for online trace selector \\
$\alpha$        & \texttt{1e-4}     & Learning rate for \texttt{Jade} \\
$\gamma'$    & \texttt{0.99} & Discounted factor \\
$\lambda$       & $log(|A|)$~\cite{li2020suphx} & Initial entropy weight \\
$\epsilon$ & \texttt{0.2}~\cite{schulman2017proximal} & PPO clip parameter \\
$c$ & \texttt{3}~\cite{ye2019mastering} & Dual-PPO clip parameter \\
$H_{target}$ & \texttt{0.1}~\cite{li2020suphx} & Target entropy of the trained policy \\
$N_{policy}$ & \texttt{5}~\cite{cobbe2020phasic} & Training iteration for the PPO \\
Optimizer & \texttt{Adam}~\cite{kingma2014adam} & Optimization method for \texttt{Jade} \\
\hline
\end{tabular}
\label{tbl:2}
\end{small}
\end{table}

\begin{table}
\caption{Datasets and metrics during testing}
\begin{small}
\begin{tabular}{ccc}
\hline
Type & Dataset/Metric & Description                 \\ \hline
Video set        &  Envivio~\cite{envivio2016} & \{0.3,~0.75,~1.2,~1.85,~2.85,~4.3\} Mbps \\ \hline
Network trace        &  HSDPA~\cite{riiser2013commute} & Slow-network paths \\
Network trace        &  FCC~\cite{bworld} & Slow-network paths \\
Network trace        &  Oboe~\cite{akhtar2018oboe}  & Medium-network paths \\
Network trace        &  FCC-18~\cite{meng2021practically} & Fast-network paths \\ \hline
QoE & QoE$_\texttt{lin}$ & The general QoE metric~\cite{pereira2018cross, yin2015control, mao2017neural,kan2022improving} \\
QoE & QoE$_\texttt{DNN}$ & DNN-based QoE metric \\
\hline
\end{tabular}
\label{tbl:3}
\end{small}
\end{table}

\end{document}